\documentclass
[aps,prc,twocolumn,showpacs,showkeys,amsmath,floatfix,superscriptaddress]{revtex4}
\usepackage{color,graphicx,ulem}
\usepackage{mathptmx}                
\usepackage{dcolumn}                 
\usepackage{bm}                      
\def\roughly#1{\mathrel{\raise.3ex\hbox{$#1$\kern-.75em%
\lower1ex\hbox{$\sim$}}}}
\def\lsim{\roughly<}
\def\gsim{\roughly>}

\begin{document}
\title{Hyperons in neutron star matter within
relativistic mean-field models}

\author{M. Oertel}
\affiliation{LUTH, CNRS, Observatoire de Paris, Universite Paris Diderot, 
5 place Jules Janssen, 92195 Meudon, France}
\author{C. Provid\^encia}
\affiliation{Centro de F\'{\i}sica Computacional, Department of Physics,
  University of Coimbra, P-3004-056 Coimbra, Portugal}
\author{F. Gulminelli}
\affiliation{ENSICAEN, UMR6534, LPC, 14050 Caen C\'edex, France}
\author{Ad. R. Raduta}
\affiliation{IFIN-HH, Bucharest-Magurele, POB-MG6, Romania}

\begin{abstract}
  Since the discovery of neutron stars with masses around $2 M_\odot$ the
  composition of matter in the central part of these massive stars has been
  intensively discussed. Within this paper we will (re)investigate the
  question of the appearance of hyperons. To that end we will perform an
  extensive parameter study within relativistic mean field models. We will
  show that it is possible to obtain high mass neutron stars (i) with a
  substantial amount of hyperons, (ii) radii of 12-13 km for the canonical mass
  of $1.4 M_\odot$, and (iii) a spinodal instability at the onset of
  hyperons. The results depend strongly on the interaction in the
  hyperon-hyperon channels, on which only very little information is available
  from terrestrial experiments up to now.
\end{abstract}

\date{\today}


\maketitle

\section{Introduction}
With the purpose of better understanding the dynamics of core-collapse
supernova and the observed neutron star characteristics, a considerable
theoretical effort has been undertaken in recent years concerning the
modelization of the equation of state (EoS) of cold dense
matter with some extensions to finite
temperature. 

If it is well admitted that hyperonic and deconfined quark matter could exist
in the inner core of neutron stars, a complete understanding of its
composition is far from being achieved.  Concerning hyperons, simple energetic
considerations suggest that they should be present at high
density~\cite{glendenning}.  However, in the standard picture, the opening
of hyperon degrees of freedom leads to a considerable softening of the
EoS~\cite{glendenning}, which in turns leads to maximum neutron star masses
smaller than the highest observed values~\cite{Demorest10, Antoniadis13}. This
puzzling situation could be circumvented by a very early deconfinement
transition~\cite{Zdunik_aa13}, or by the population of other baryonic states
such as $\Delta$-baryons pushing the hyperon onset to higher densities. The
latter, however, only replaces the hyperon puzzle by a $\Delta$
puzzle~\cite{Drago2014} unless a phase transition to quark matter is simultaneously invoked.

Without calling upon such a transition, it has been shown that the
observed neutron star masses simply imply that the hyperon-hyperon ($YY$) and
hyperon-nucleon ($YN$) couplings must be much more repulsive at high density
than presently assumed (e.g.~\cite{Hofmann00,Stone07,Bednarek2012,
  weissenborn11,Weissenborn2012,sedrakian_aa2012,micaela_prc85,
  Colucci2013,Lopes2014, Banik2014, vanDalen2014, Katayama2014, Gomes2014}). The general agreement is,
however, that the price to pay for this additional repulsion is a very low
strangeness content of neutron stars~\cite{Weissenborn2012}. In
Ref.~\cite{Raduta2014} it was shown, that at least in non-relativistic models,
this is strongly parameter dependent and that neutron stars with a
considerable amount of hyperons can exist.

Another argument put forward against hyperons is that the strong
repulsion needed, has its implications in the purely nuclear part,
too, leading to large radii for neutron stars, larger than those
suggested for intermediate mass neutron star by recent
observations~\cite{Fortin2014}. In many models with hyperons
compatible with the neutron star mass constraint, see
e.g.~\cite{Bednarek2012, Weissenborn2012}, indeed relatively high
radii of about 14 km for a non-rotating spherical neutron star with
the canonical mass of $1.4 M_\odot$ are obtained. This can, however,
not be a general argument since there are some examples with lower
radii ~\cite{Stone07,Colucci2013, Banik2014}.

In addition, the generic presence of attractive and repulsive couplings
suggests the existence, in a model-independent manner, of a phase transition
involving strangeness.  A detailed study of the phase diagram of dense
baryonic matter was recently undertaken in Refs.~\cite{nL,npLe,Raduta2014}
within a non-relativistic mean-field model based on phenomenological
functionals. It was shown that under these assumptions first- and second-
order phase transition exist, and are expected to be explored under the
strangeness equilibrium condition characteristic of stellar matter. In
Ref.~\cite{Schaffner2000} such a phase transition has been discussed for
relativistic mean field models, but within a model with very strong $YY$
attraction.

Here we are interested (a) in examining in which region of parameter space
within RMF models a first order phase transition from purely nuclear to
hyperonic matter could exist and if the existence of such a phase transition
is compatible with experimental and observational data, (b) if it is possible
to obtain high mass neutron stars with considerable amount of hyperons and (c)
in finding additional support for hyperonic EoS with radii of 12-13 km for
canonical $1.4 M_\odot$ neutron stars and maximum masses in agreement with
observations.

We will work at zero temperature throughout the whole paper. The paper is
organized as follows. In Sec.~\ref{sec:model} we discuss the model applied and
the setup for the hyperonic interactions. We explain the procedure to detect
thermodynamic instabilities in Sec.~\ref{sec:thermo}. In Sec.~\ref{sec:lambda}
we present results for neutron star matter containing nucleons, electrons and
$\Lambda$ hyperons. We extent the discussion to the full baryonic octet in
Sec.~\ref{sec:fulltov} and we close the paper by a summary of results in
Sec.~\ref{sec:summary}.

\section{The model}
\label{sec:model}
The literature on phenomenological RMF models including hyperons is large and
many different versions exist (see e.g.~\cite{Dutra2014}), including either
non-linear couplings or density-dependent ones of baryons to the meson fields
mediating the interaction. Let us stress at this point that, although they are
generally called meson fields, these fields are purely phenomenological and
only serve for describing the interaction without any relation with existing
meson fields, except for the quantum numbers which give the names for the
corresponding RMF meson fields.  The Lagrangian of the model can be written in
the following form
\begin{eqnarray}
{\mathcal L} &=& \sum_{j \in \mathcal{B}}  \bar \psi_j \left( i \gamma_\mu \partial^\mu - m_j +
  g_{\sigma j} \sigma
  + g_{\sigma^* j} \sigma^* \right.\nonumber \\ &&\left. + g_{\delta
    j} \vec{\delta} \cdot \vec{I}_j - g_{\omega j} \gamma_\mu \omega^\mu - g_{\phi j}
\gamma_\mu \phi^\mu - g_{\rho j} \gamma_\mu \vec{\rho}^\mu \cdot \vec{I}_j\right) \psi_j \nonumber \\ &&
 + \frac{1}{2} (\partial_\mu \sigma \partial^\mu \sigma - m_\sigma^2 \sigma^2)
  - \frac{1}{3} g_2 \sigma^3 - \frac{1}{4} g_3 \sigma^4 \nonumber \\ && 
 + \frac{1}{2} (\partial_\mu \sigma^* \partial^\mu \sigma^* - m_{\sigma^*}^2
  {\sigma^*}^2) \nonumber \\ &&
 + \frac{1}{2} (\partial_\mu \vec{\delta} \partial^\mu \vec{\delta} - m_{\delta}^2
  {\vec{\delta}}^2) \nonumber \\ &&
- \frac{1}{4}
W^\dagger_{\mu\nu} W^{\mu\nu} 
- \frac{1}{4}
P^\dagger_{\mu\nu} P^{\mu\nu} 
- \frac{1}{4}
\vec{R}^\dagger_{\mu\nu} \vec{R}^{\mu\nu} \nonumber \\ && 
+ \frac{1}{2} m^2_\omega \omega_\mu \omega^\mu + \frac{1}{4} c_3 (\omega_\mu \omega^\mu)^2 \nonumber \\ && 
+ \frac{1}{2} m^2_\phi \phi_\mu \phi^\mu 
+ \frac{1}{2} m^2_\rho \vec{\rho}_\mu \vec{\rho}^\mu ~,
\end{eqnarray}
where $\psi_j$ denotes the field of baryon $j$, and $W_{\mu\nu}, P_{\mu\nu}, \vec{R}_{\mu\nu}$ are the
vector meson field tensors of the form
\begin{eqnarray}
V^{\mu\nu} = \partial^\mu  V^\nu - \partial^\nu V^\mu~.
\end{eqnarray}
$\sigma, \sigma^*$ are scalar-isoscalar  meson
fields, coupling to all baryons ($\sigma$) and to strange baryons
($\sigma^*$), respectively. $\vec{\delta}$ induces a scalar-isovector
coupling. 

In the mean field approximation, the meson fields are
replaced by their respective mean-field expectation values, which are given in
uniform matter as
\begin{eqnarray}
m_\sigma^2 \bar\sigma + g_2 \bar\sigma^2 + g_3 \bar\sigma^3 &=& \sum_{i \in B} g_{\sigma i} n_i^s
\\
m_{\sigma^*}^2 \bar\sigma^* &=& \sum_{i \in B} g_{\sigma^* i} n_i^s\\
m_\delta^2 \bar\delta &=& \sum_{i \in B} g_{\delta i} t_{3 i} n_i^s\\
m_\omega^2 \bar\omega + c_3 \bar\omega^3 &=& \sum_{i \in B} g_{\omega i} n_i\\
m_\phi^2 \bar\phi &=& \sum_{i \in B} g_{\phi i} n_i\\
m_\rho^2 \bar\rho &=& \sum_{i \in B} g_{\rho i} t_{3 i} n_i~,
\end{eqnarray}
where  $\bar\delta=\langle\delta_3\rangle$, $\bar\rho=\langle\rho_3^0\rangle$,
$\bar\omega=\langle\omega^0\rangle$, $\bar \phi=\langle\phi^0\rangle$, and
$t_{3 i}$ represents the third component of isospin of baryon $i$ with the convention that
$t_{3 p} = 1/2$. The scalar density of baryon $i$ is given by
\begin{equation}
n^s_i = \langle \bar \psi_i \psi_i \rangle = \frac{1}{\pi^2} \int_0^{k_{Fi}}
k^2 \frac{M^*_i} {\sqrt{k^2 + M^{*2}}} dk~,
\end{equation}
and the number density by
\begin{equation}
n_i = \langle \bar \psi_i\gamma^0 \psi_i \rangle = \frac{1}{\pi^2} \int_0^{k_{Fi}}
k^2 dk = \frac{k_{Fi}^3}{3 \pi^2}~.
\end{equation}
The effective baryon mass $M^*_i$ depends on the scalar mean fields as
\begin{equation}
M^*_i = M_i - g_{\sigma i} \bar\sigma - g_{\sigma^* i} \bar\sigma^* -
g_{\delta i} t_{3 i} \bar \delta~,
\end{equation}
and the effective chemical potentials, $(\mu_i^*)^2 = (M_i^*)^2 + k_{Fi}^2$,
are related to the chemical potentials via
\begin{equation}
\mu_i^* = \mu_i - g_{\omega i} \bar\omega - g_{\rho i} \,t_{3 i}
\bar\rho - g_{\phi i} \bar \phi - \Sigma_0^R~.
\label{mui}
\end{equation} 
The rearrangement term 
\begin{eqnarray}
\Sigma_0^R &=& \sum_{j \in B} \left( \frac{\partial g_{\omega j}}{\partial n_j}
\bar\omega n_j + t_{3 j} \frac{\partial g_{\rho j}}{\partial n_j}
\bar\rho n_j +\frac{\partial g_{\phi j}}{\partial n_j}
\bar\phi n_j \right. \nonumber \\ && \left. -\frac{\partial g_{\sigma j}}{\partial n_j}
\bar\sigma n_j^s -\frac{\partial g_{\sigma^* j}}{\partial n_j}
\bar\sigma^* n_j^s - t_{3 j} \frac{\partial g_{\delta j}}{\partial n_j}
\bar\delta n_j^s \right)~.
\end{eqnarray}
is present in density dependent models to ensure thermodynamic consistency. 

For the present study we will limit ourselves to two non-linear models,
GM1~\cite{gm91} and TM1-2~\cite{providencia13}, and one density-dependent
model, DDH$\delta$~\cite{Avancini2009,Gaitanos}. The two non-linear ones
have been chosen among the large number of models since
they are two widely used ones, with a very different strategy for determining
the parameters: GM1 has been adjusted to nuclear saturation properties
imposing a certain effective mass and incompressibility, whereas TM1 has been
fitted to ground state properties of nuclei and at high densities to DBHF
calculations.  For the GM1 parametrization, $c_3$ =0, and the $\delta$-field
is absent in GM1 and TM1-2. The density-dependent models assume $g_2 = g_3 =
c_3 = 0$ (no non-linear terms) and the following density dependence of the
couplings is used within DDH$\delta$
\begin{equation}
g_i(n_B) = g_i(n_0) h_i(x)~,\quad x = n_B/n_0~,
\end{equation}
with $n_0$ denoting nuclear matter saturation density and 
\begin{equation}
h_i(x) = a_i \frac{1 + b_i ( x + d_i)^2}{1 + c_i (x + d_i)^2}
\end{equation}
for all isoscalar couplings and 
\begin{equation}
h_\rho(x) = a_i\,\exp[-b_i (x-1)] - c_i (x-d_i)~.
\end{equation}
for the isovector ones.  The parameter values for GM1 can be found,
e.g., in Ref.~\cite{Miyatsu2013}, Table III, for TM1-2 in
Ref.~\cite{providencia13}, table I, and for the DDH$\delta$ model in
Ref.~\cite{Avancini2009}, Table II.  The EoS of homogeneous symmetric
nuclear matter for parametrization TM1-2 is shown in Fig.~1 of
Ref.~\cite{providencia13}. It has the same properties as TM1
\cite{tm1} at and below saturation density, but it is stiffer at
supra-saturation densities, still within the constraints imposed by
heavy-ion flow~\cite{danielewicz2002}. However, within this
parametrization the slope of the symmetry energy at saturation is very
large, $L=110$ MeV.  Since the radius of compact stars is very
sensitive to $L$~\cite{hor01,hor03,rafael11}, we will consider a
modified version with $L=55$ MeV~\cite{providencia13}, too,
introducing a non-linear $\omega-\rho$ term as in Ref.~\cite{hor01}.
The resulting properties of homogeneous symmetric nuclear matter are
listed in Table~\ref{tab:nuclear} for all parametrizations employed
within the present paper. In the same table we also include the value
of the pressure of $\beta$-equilibrated cold neutron star matter at
$n_0$ for reference.  According to Ref.~\cite{hebeler2013}, where a
microscopic neutron matter calculation in the framework of a chiral
effective field theory together with available information on
symmetric nuclear matter have been used to build the EoS of stellar
matter, this value should lie in the range
\begin{equation}
1.8 \lsim P(n_0)\lsim 3.0 \mbox{ MeV/fm}^3.
\label{hebeler}
\end{equation}
This range of pressures is the result of a quite restrictive allowed
region for the symmetry energy (29.7-33.5 MeV) and its slope $L$
(32.4-57 MeV) at saturation.

\begin{table}[h!]
 \begin{tabular}{c|cccccc}
  &$K$& $E_\mathit{sym}$ & $n_0 $& $B$ 
  &$L$  &$P(n_0)$\\ 
& [MeV] & [MeV] & [MeV] & $[\mathrm{fm}^{-3}]$& [ MeV] & [ MeV/fm$^3$] \\
\hline \hline
GM1 &300&32.5&0.153&16.3&94 &4.06\\
TM1-2&281&36.9&0.145&16.3&110/55& 4.38/2.43 \\
DDH$\delta$ &240&25.1 &0.153&16.3&44 & 2.56\\
\end{tabular}
\caption{\it Nuclear matter properties of the models considered in
  this study for symmetric nuclear matter at saturation, except for the last
  column where  the pressure of stellar matter at $n_0$
is given}
\label{tab:nuclear}
\end{table}

\subsection{Setup for the hyperonic interaction}
\label{sec:hyperons}
The wealth of nuclear data allows to constrain the nuclear interaction
parameters within reasonable ranges, whereas this is not the case for
hyperons, where data are scarce. These leaves some freedom in adjusting the
interaction parameters for the hyperonic sector. 

Many recent works, see e.g.~\cite{Banik2014,Miyatsu2013,Weissenborn2012}, use
a procedure inspired by the symmetries of the baryon octet to
express the individual isoscalar vector meson-baryon couplings in terms of
$g_{\omega N}$ and a few additional parameters~\cite{Schaffner1996} as follows
{\small
\begin{eqnarray}
\frac{g_{\omega\Lambda}}{g_{\omega N}} &=& \frac{ 1 - \frac{2 z}{\sqrt{3}} 
  (1-\alpha) \tan \theta}{1- \frac{z}{\sqrt{3}}  (1 - 4 \alpha)
  \tan\theta} ~,\  
\frac{g_{\phi\Lambda}}{g_{\omega N}} = -\frac{ \tan\theta + \frac{2 z}{\sqrt{3}} 
  (1-\alpha)}{1- \frac{z}{\sqrt{3}}  (1 - 4 \alpha)
  \tan\theta} ~,\nonumber \\
\frac{g_{\omega\Xi}}{g_{\omega N}} &=& \frac{ 1 - \frac{z}{\sqrt{3}} 
  (1+ 2 \alpha) \tan \theta}{1- \frac{z}{\sqrt{3}}  (1 - 4 \alpha)
  \tan\theta} ~,\ 
\frac{g_{\phi\Xi}}{g_{\omega N}} = -\frac{ \tan\theta + \frac{z}{\sqrt{3}} 
  (1+ 2 \alpha)}{1- \frac{z}{\sqrt{3}}  (1 - 4 \alpha)
  \tan\theta} ~,\nonumber \\
\frac{g_{\omega\Sigma}}{g_{\omega N}} &=& \frac{ 1 + \frac{2 z}{\sqrt{3}} 
  (1- \alpha) \tan \theta}{1- \frac{z}{\sqrt{3}}  (1 - 4 \alpha)
  \tan\theta} ~,\ 
\frac{g_{\phi\Sigma}}{g_{\omega N}} = \frac{ -\tan\theta + \frac{ 2 z}{\sqrt{3}} 
  (1- \alpha)}{1- \frac{z}{\sqrt{3}}  (1 - 4 \alpha)
  \tan\theta} ~,\nonumber \\
\frac{g_{\phi N}}{g_{\omega N}} &=& -\frac{ \tan\theta + \frac{z}{\sqrt{3}} 
  (1- 4 \alpha)}{1- \frac{z}{\sqrt{3}} (1 - 4 \alpha)
  \tan\theta} ~.
\label{eq:symmetry}
\end{eqnarray} 
}
The parameter $\alpha$ thereby determines the ratio of symmetric coupling of
the baryons to the vector meson octet ($D$-term) and the antisymmetric
coupling ($F$-term), and $g_1$ and $g_8$ are the coupling constants for
coupling of baryons to the vector meson singlet and octet,
respectively. $\theta$ is the mixing angle of $\omega$- and $\phi$-mesons with
the corresponding singlet and octet states, and $z = g_8/g_1$.  As commonly
assumed, in what follows, we will take $\tan\theta = 1/\sqrt{2}$,
corresponding to ideal mixing and $\alpha = 1$. In the literature, it
is mostly imposed
$SU(6)$-symmetry to fix the couplings, i.e. $z = 1/\sqrt{6}$, and
only recent studies in view of the observation of high mass neutron stars have
relaxed this assumption, for
example~\cite{Weissenborn2012,Miyatsu2013,Lopes2014}.

In the isovector sector, not the same procedure is applied, since this would
lead to contradictions with the observed nuclear symmetry energy. $g_{\rho N}$
is therefore left as a free parameter, adjusted to the desired value of the
symmetry energy, and the remaining nonvanishing isovector
vector couplings are all equal, the isospin symmetry being taken into
  account through the isospin operator $t_{3i}$, see Eq. (\ref{mui}),
\begin{equation}
g_{\rho \Lambda} = 0,\quad g_{\rho N} = g_{\rho \Xi} = g_{\rho\Sigma}~.
\end{equation}

For the scalar sector, in Ref.~\cite{Colucci2013} a symmetry inspired
procedure is discussed tested against the constraints imposed by hypernuclear
data. Here we will use directly, as done e.g. in Refs.~\cite{Weissenborn2012,
  Banik2014}, the information from hypernuclear data on hyperonic
single-particle mean field potentials to constrain the coupling constants.
The potential for particle $j$ in $k$-particle matter is given by
\begin{equation}
U_j^{(k)}(n_k) = M^*_j - M_j + \mu_j - \mu^*_j~.
\end{equation}
Based on data on $\Lambda$-hypernuclei produced in $(\pi^+,K^+)$ reactions,
the presently accepted value of the $\Lambda$-potential in symmetric nuclear
matter at saturation density, $U_{\Lambda}^{(N)}(n_0)$, is $ \approx -30$
MeV~\cite{UNL}.  $U_{\Xi}^{(N)}(n_0)$ is attractive, too, with a value of
$\approx -14$-$-18$ MeV, based on missing mass measurements in the $(K^-,K^+)$
reaction on carbon~\cite{UNXi}.  The situation of $U_{\Sigma}^{(N)}(n_0)$ is
ambiguous.  On the one hand $(\pi^-,K^+)$ reactions on medium-to-heavy nuclei
point to a repulsive potential of up to 100 MeV~\cite{UNSigma_repulsive}.  On
the other hand, the observation of a $^4_{\Sigma}$He bound state in a
$^4$He($K^-,\pi^-$) reaction~\cite{UNSigma_attractive} pleads in favor of an
attractive potential.  Following the above procedure to fix the vector
coupling constants, the couplings of hyperons to $\sigma$ are then adjusted to
reproduce the hyperon potentials
in symmetric nuclear matter. 

Very few multi-hyperon exotic nuclei data exist so far and all of them
correspond to double-$\Lambda$ light nuclei. Data on the bond energy
can be reinterpreted in terms of the $\Lambda$ potential in $\Lambda$
matter at the average density of $\Lambda$ inside those
nuclei~\cite{vidana_2001}.  Mean-field calculations suggest that in
light nuclei (from He to C) the average $\Lambda$ density is close to
one fifth of the saturation density \cite{khan,vidana_2001}.Therefore, we take as
an indicative value for $U_\Lambda^{(\Lambda)} (n_0/5)$ the experimental value
of $\Delta B_{\Lambda}^{\Lambda}$. 
Data on $^{10}_{\Lambda \Lambda}$Be
and $^{13}_{\Lambda \Lambda}$B then suggest
$U_\Lambda^{(\Lambda)}(n_0/5) \approx -5$ MeV \cite{exp_2lambda} while
$^{6}_{\Lambda \Lambda}$He data point toward a higher value of
$U_\Lambda^{(\Lambda)}(n_0/5)\approx - 0.67$ MeV \cite{Aoki_2009,ahn_2013}.

We will use here as a guideline that experimental data point towards
$U_\Lambda^{(\Lambda)}(n_0/5) > -5 $ MeV. For the other potentials it
is often assumed that in isospin symmetric $\Xi$- and $\Sigma$-matter,
$U_{\Xi}^{(\Xi)}(n_0) \approx 2 U_\Lambda^{(\Lambda)}(n_0/2)$ and $
U_{\Sigma}^{(\Sigma)}(n_0) \approx
U_\Lambda^{(\Lambda)}(n_0/2)$~\cite{Schaffner1994} based on
theoretical estimates. In view of the only weakly attractive
$\Lambda\Lambda$-potential and the uncertainties on other
hyperon-hyperon ($YY$) potentials, often $\sigma^*$ is neglected (see
e.g.~\cite{Banik2014,Weissenborn2012,Bednarek2012}).

In a first step we will use the procedure described above and study the
dependence of the results upon variations of the couplings to $\sigma^*$ and
$z$. The symmetry arguments in the isoscalar vector sector are, however, not
very compelling. They are based on the naive quark model for hadrons in vacuum
and it is known that this model is too simple. The interactions of the baryon
octet in vacuum respect an approximate $SU(3)$-flavor symmetry, but symmetry
breaking effects are large. Since we are dealing here with an effective
model for interacting particles in matter, without any input about symmetry
breaking effects in dense matter, there is no reason to assume any flavor
$SU(3)$-symmetry for the effective interaction. In addition, the approach is
inconsistent in the sense that symmetry constraints are imposed only for the
isoscalar vector couplings with other prescriptions for the other channels, see
also the discussion in Ref.~\cite{Lopes2014} on this point. By the way, in the
vector-isovector channel, a strict application of this procedure would lead to
severe problems with the observed nuclear symmetry
energy~\cite{Weissenborn2012}. Therefore, in a second step, we will only keep
the $NY$- potentials at some given value and vary the different
coupling parameters freely.

\section{Thermodynamic instabilities}
\label{sec:thermo}
The existence of a first order phase transition can be spotted by analyzing
the curvature of a thermodynamic potential in terms of extensive variables,
indicating the presence of a spinodal instability related to the phase
transition. The unstable region is thereby recognized by a negative curvature.
This convexity analysis has been often employed, among others for the phase
transition to hyperonic matter in Refs.~\cite{nL,npLe,Raduta2014} or for the
neutron-proton system~\cite{Avancini06,ducoin_2006,Ducoin08}. At zero
temperature, the adequate thermodynamic potential is given by the total energy
density, $\varepsilon (n_i)$, with as variables the number densities
corresponding to good quantum numbers.

In the present case, assuming equilibrium with respect to strong and
electromagnetic interaction, for purely baryonic matter the good quantum
numbers are baryon number, strangeness and charge with densities $n_B, n_S$,
and $n_Q$, respectively. Since we are interested in neutron star matter, we
have to impose electrical charge neutrality and add leptonic degrees of
freedom with lepton number as an additional degree of freedom. Due to the
strict electrical neutrality condition, charge is no longer a good degree of
freedom and the system remains three-dimensional~\cite{anomalous,npLe}, see
also Ref.~\cite{providencia_electrons}, in terms of the number densities $n_B,
n_S$ and $n_L$~\footnote{Let us stress that for the stability analysis we
  assume that the time-scale of potential density fluctuations is such that
  equilibrium with respect to strong interaction is always maintained. This
  allows to reduce the nine dimensional space of all densities to a three
  dimensional one. In this sense our study corresponds to the semi-frozen case
  of Ref.~\cite{Gusakov2014}.}

Stability can now be checked by analyzing the eigen-values of the curvature
matrix, $C_{ij}=
\partial^2 \varepsilon(\{n_l\}_{l=\{i,j,k\}})/\partial n_i \partial n_j $,
where $i,j,k=B,S,L$.  The number of negative eigenvalues corresponds to the
number of directions in density space, in which density fluctuations get
spontaneously and exponentially amplified in order to achieve phase
separation. In all our studies at most one negative eigenvalue has been found.

Muons could be included into the analysis, since they are a priori present in
neutron star matter. Neglecting neutrino oscillations, they would add another
dimension, corresponding to conserved muon lepton number. However, as leptons
are treated as an ideal gas, they change the stability analysis only through
the electrical charge neutrality constraint. Therefore, including muons in
addition to electrons does not qualitatively change the results. The
quantitative modifications are so small that we have decided to neglect muons
for simplicity.

To explore the complete three-dimensional space $n_B, n_S, n_L$ is a very
demanding task. Since we are mainly interested in neutron star matter, we will
restrict our investigation to the case of strangeness changing weak
equilibrium, i.e. $\mu_S = 0$. In addition we will assume $\beta$-equilibrium.
We will thereby consider that a neutron star older than several minutes is cold
enough such that neutrinos can freely leave the system. This means that their
chemical potential is zero, i.e. the chemical potentials associated with
(electron) lepton number vanishes, $\mu_L = 0$.

Let us stress that, although we restrict the study to a line in the
three-dimensional density space, the stability analysis remains
three-dimensional: at every point on the $\mu_S = 0, \mu_L = 0$-line, the
curvature matrix tests fluctuations in three directions, meaning that we do
not assume weak equilibrium to be maintained throughout the fluctuations, see
Ref.~\cite{Gusakov2014}, too. 

\section{Results and discussion}
The results presented below assume $U_{\Lambda}^{(N)}(n_0) = -28$ MeV,
$U_{\Xi}^{(N)}(n_0) = -18$ MeV, and $U_{\Sigma}^{(N)}(n_0) = 30$ MeV unless
otherwise stated. To study
the parameter dependence, we will vary the different coupling constants
keeping the nuclear matter properties of the different models constant. 
\subsection{Nuclear matter with $\Lambda$-hyperons}
\label{sec:lambda}
\subsubsection{Stability analysis}
\label{sec:lambdastable}
We will start the discussion with the simple case of nuclear matter with
$\Lambda$-hyperons. Although not completely realistic, it is instructive since
it allows to see trends in the parameter dependence. As mentioned earlier, in
a first step we will follow the procedure proposed in
Ref.~\cite{Weissenborn2012} to vary the coupling constants, see
Section~\ref{sec:hyperons} for details. The values of the isoscalar vector
couplings are then determined by the value of the parameter $z$ and the
couplings to $\sigma$ are obtained from the hyperon potentials in nuclear
matter. We will thereby vary $z$ between $z = 0$ and $z = 1/\sqrt{6}$, the
$SU(6)$ value, since for higher values of $z$ it becomes more and more
difficult to obtain neutron star maximum masses in agreement with
observations, see Ref.~\cite{Weissenborn2012} and
Section~\ref{sec:lambdaneutronstar}.

\begin{figure*}
\begin{center}
\includegraphics[angle=0, width=0.48\textwidth]{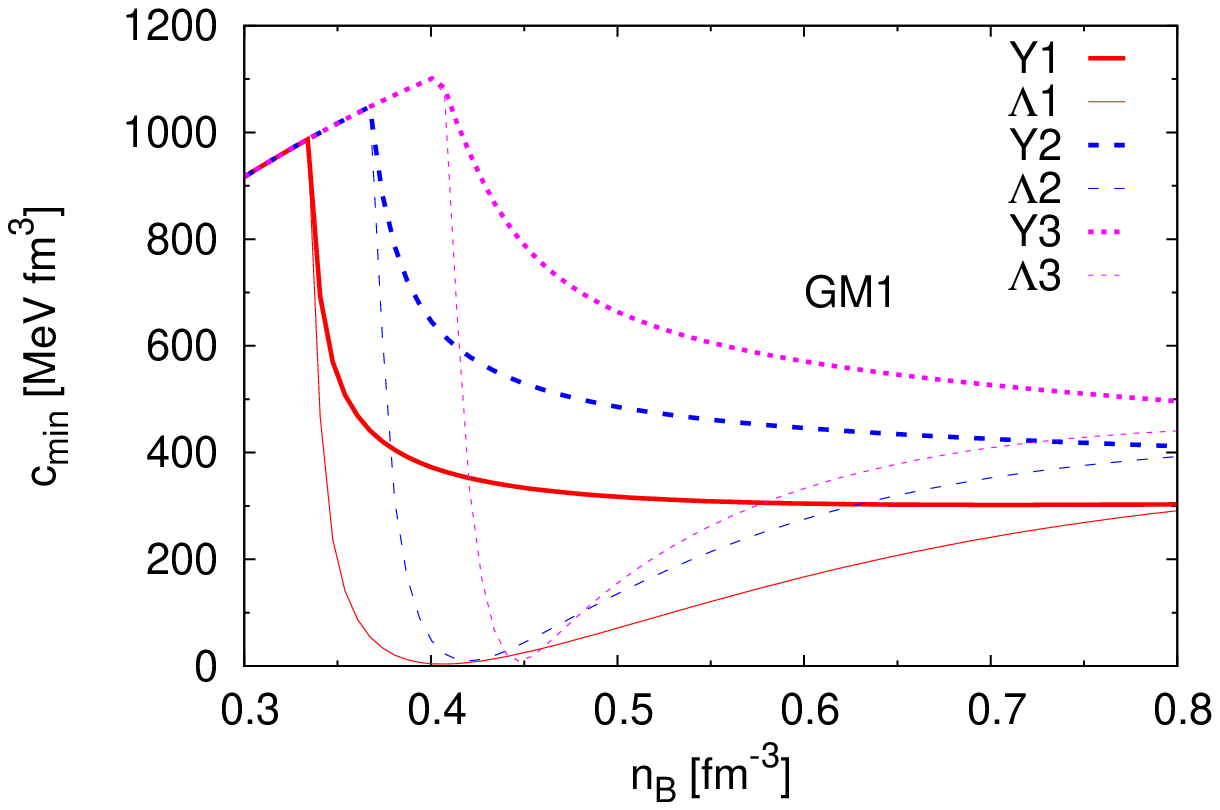}\hfill
\includegraphics[angle=0, width=0.48\textwidth]{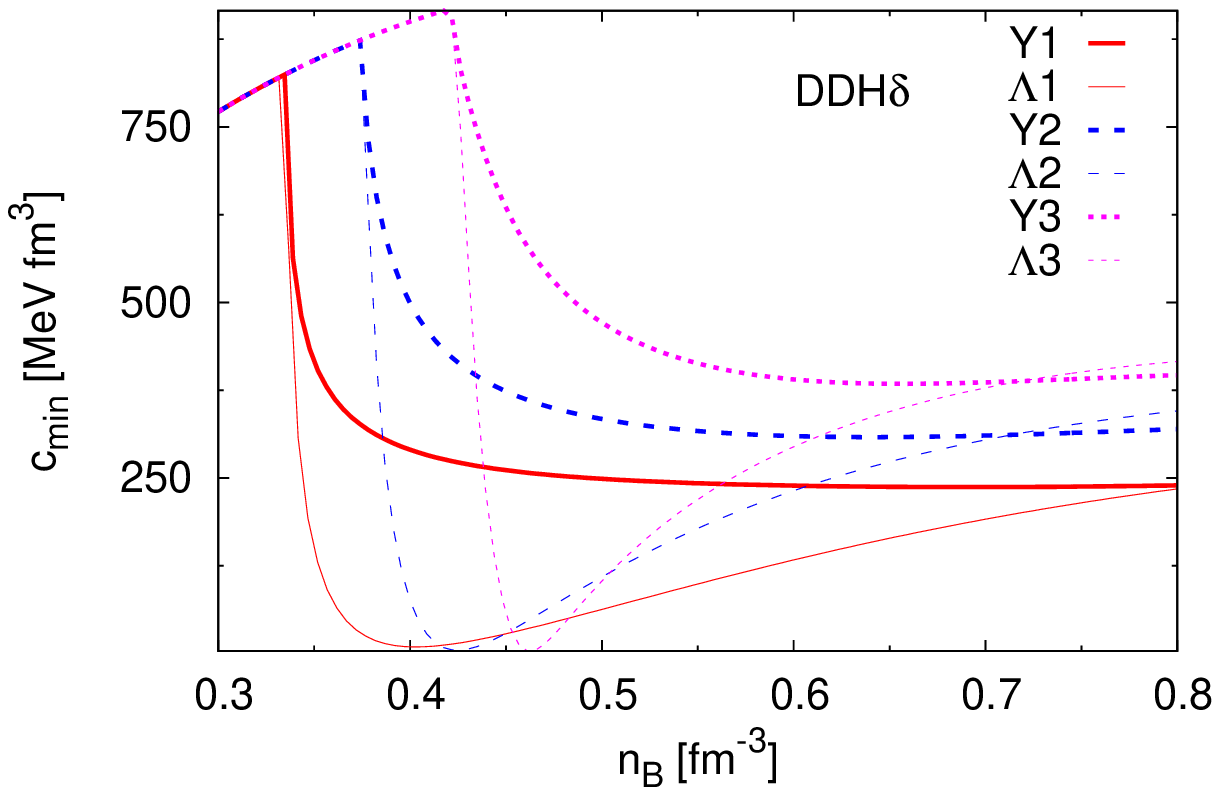}
\end{center}
\caption{\it Smallest eigenvalue of the curvature matrix of the energy density
  as a function of baryon number density for neutron star matter with
  $R_{\sigma^* i} = g_{\sigma^* i}/g_{\sigma N}$. The GM1 and DDH$\delta$
  parameter sets have been employed. Only $\Lambda$ hyperons are
  considered. To test the parameter dependence we have respected the symmetry
  constraints in the isoscalar vector sector, see
  Eq.~(\ref{eq:symmetry}). The couplings to $\sigma$ are adjusted to the
  hyperon potentials in nuclear matter and the $\sigma^*$ are varied,
  see Section~\ref{sec:hyperons} for more details.}
\label{fig:lambdasymcritical}
\end{figure*}

In Fig.~\ref{fig:lambdasymcritical} the minimal eigenvalue of the curvature
matrix, $c_{\mathit{min}}$, is displayed as a function of baryon density for
neutron star matter within GM1 and DDH$\delta$-models, varying $z$ and
$g_{\sigma^* \Lambda}$. The kink in the curves indicates the respective
threshold density for the onset of $\Lambda$-hyperons. The lower value of
$g_{\sigma^* \Lambda}$ has thereby been chosen such that the value
$U_{\Lambda}^{(\Lambda)} (n_0/5) = -5$ MeV is reproduced, which corresponds to
the strongest attraction compatible with present experimental data~\footnote{
  Please note that for vanishing $g_{\sigma^* \Lambda}$, as always assumed in
  the recent literature, in most cases no attraction at all in the
  $\Lambda$-potential is observed.}. Although, after the onset of
$\Lambda$-hyperons, $c_{\mathit{min}}$ decreases with increasing $z$, no
instability is found in this case. Upon increasing $g_{\sigma^* \Lambda}$,
$c_{\mathit{min}}$ further decreases and for $g_{\sigma^* \Lambda} >
g_{\mathit{crit}}$ a first order phase transition in neutron star matter can
be observed.

\begin{figure*}
\begin{center}
\includegraphics[angle=0, width=0.48\textwidth]{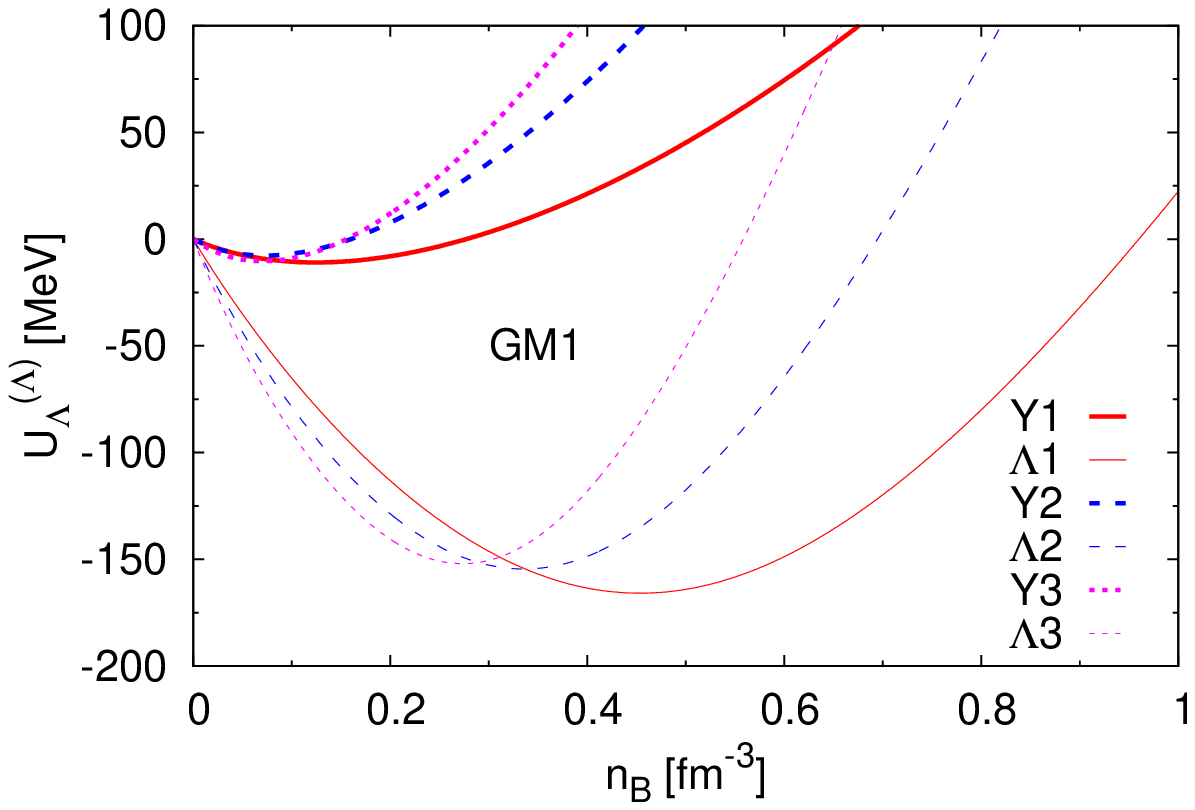}\hfill
\includegraphics[angle=0, width=0.48\textwidth]{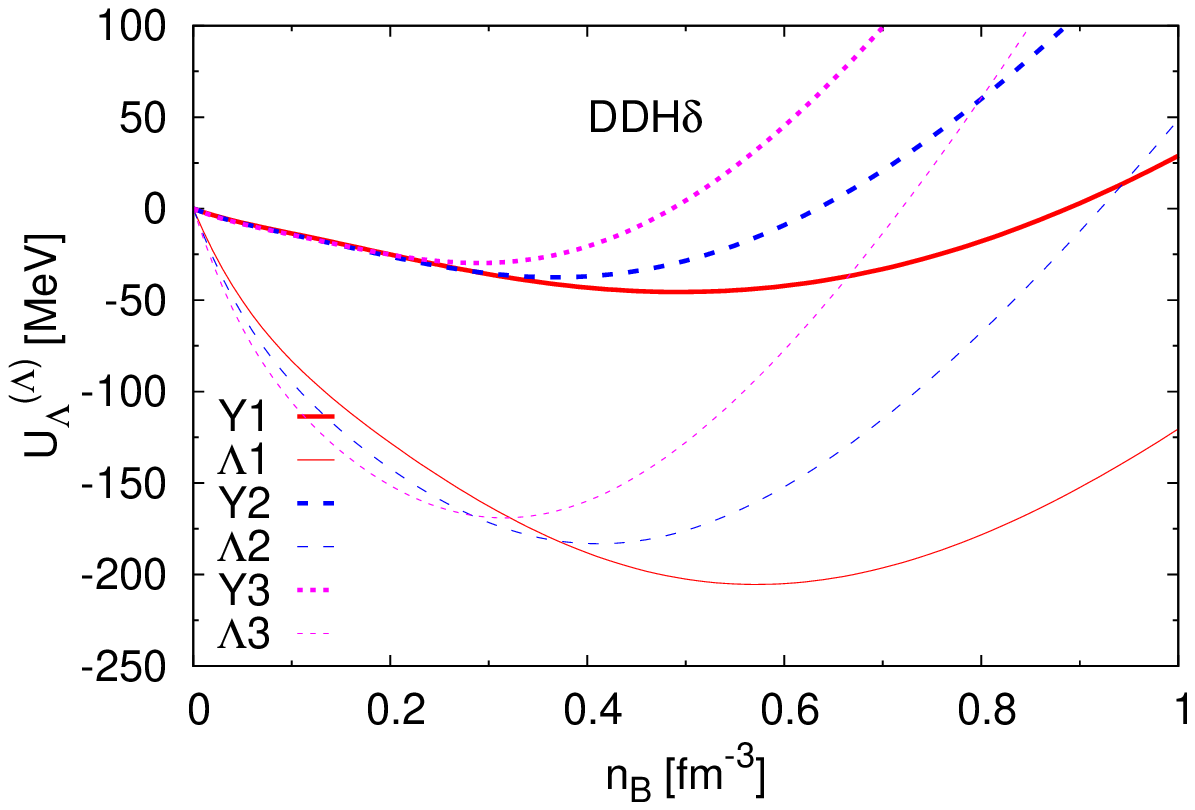}
\end{center}
\caption{\it The $\Lambda$ single particle potential in $\Lambda$-matter as
  function of baryon density. The GM1 (left) and
  DDH$\delta$ (right) 
  parameter sets have been employed. The setup is the same as in
  Fig.~\ref{fig:lambdasymcritical}.}
\label{fig:lambdasymupotentials}
\end{figure*}

At the same time, increasing the value of $g_{\sigma^* \Lambda}$ renders the
$\Lambda\Lambda$ interaction more attractive at low densities. The values of
$U_\Lambda^{(\Lambda)}(n_0/5)$ are given for $g_{\sigma^* \Lambda} =
g_{\mathit{crit}}$ in Table~\ref{tab:lambdasym} for comparison and the
respective density dependence is shown in
Fig.~\ref{fig:lambdasymupotentials}. The present results suggest that a first
order phase transition to $n p \Lambda$ matter can occur in RMF models,
too. The price to pay is a very strong $\Lambda$-$\Lambda$-attraction, which
is in contradiction with the actual experimental information. In the following
we use the definition $R_{\sigma^* i} = g_{\sigma^* i}/g_{\sigma N}$.

\begin{table}[h!]
 \begin{tabular}{c||cccccc}
  Model&$R_{\sigma^* \Lambda}$& $z$ & $M_{\mathit{max}}$& $R_{1.4}$ &$R_\mathit{max}$&
  $U_{\Lambda}^{(\Lambda)}(n_0/5) $  
\\
& & &$[M_\odot]$ & [km]& [km]& [MeV] \\ \hline \hline
 GM1, $Y1$ &0.45& 0.41 & 1.99 &13.8&12.0&  -5   \\
  $\Lambda 1$&0.98 &0.41 & 1.70      &13.8 &11.1& -22 \\ \hline
  $Y2$&0.2 &0.2 & 2.22      & 13.8 &12.0&-5   \\
 $\Lambda 2$&1.02&0.2 &  2.03     & 13.8 &11.2&-28\\ \hline
 $Y3$&0&0. &  2.32     & 13.8 &12.0&-7\footnote{This value is the highest one which can be
   obtained within the model and the given parameter set.}   \\
 $\Lambda 3$&1.08&0. &  2.22     & 13.8 &11.5& -34 \\ \hline \hline
 DDH$\delta$, $Y1$  & 0.61&0.41 & 1.71   &12.7&10.4    & -5    \\
  $\Lambda 1$ &1.07 &0.41 & 1.58  &10.5& 9.0   &  -34 \\ \hline
  $Y2$ &0.5&0.2 &   1.93&12.7 &  10.8 &  -5   \\
  $\Lambda 2$ &1.08 &0.2 &  1.83   &12.7 & 9.8 &-39 \\ \hline
  $Y3$ &0.4 &0. &   2.06   &12.7&11.1 &  -5   \\
  $\Lambda 3$ &1.1 &0. &  1.99     &12.7 &10.5 & -44
\end{tabular}
\caption{\it Summary of results respecting the symmetry arguments for
  the isoscalar vector couplings allowing for $\Lambda$ hyperons as
  the only hyperons. $R_{1.4}$ denotes the radius of a non-rotating
  star with $M = 1.4 M_\odot$ and $R_\mathit{max}$ is the radius at
  maximum mass.  The values of $R_{\sigma^*\Lambda}$ for models
  $\Lambda n$ thereby correspond to the critical values of
  these coupling constants for the onset of an instability in the
  $\Lambda$ -channel, and models denoted as $Y n$ do not present any instability. 
  \label{tab:lambdasym}}
\end{table}

Relaxing the symmetry conditions, see Eq.~(\ref{eq:symmetry}), for the
variation of the isoscalar vector coupling constants leads to essentially the
same conclusion on the stability of neutron star matter with
$\Lambda$-hyperons: An instability shows up for strongly attractive
$\Lambda\Lambda$ interactions. The value of $U_{\Lambda}^{(\Lambda)} (n_0/5)$
for which the instability sets in depends only very weakly on the values of
the isoscalar vector couplings and is $U_{\Lambda}^{(\Lambda)}(n_0/5) \approx -
40$ MeV within the DDH$\delta$-model and $U_{\Lambda}^{(\Lambda)}(n_0/5)
\approx - 30$ MeV within GM1.

It is interesting to observe that these results are very different
from what is obtained in the non-relativistic framework
\cite{Raduta2014}.  In that study the parameter space associated to an
instability is very large and includes the present hypernuclear
experimental constraints. Within the present models, we arrive to the
opposite conclusion. We cannot exclude that the different functional
forms associated to the energy density in the relativistic and
non-relativistic framework might be at the origin of this discrepancy,
since it has been observed in the past that non-relativistic
functionals often present unphysical instabilities \cite{non-rel}.  An
alternative explanation might be the qualitatively different behavior
of the hyperon-hyperon potentials, which in both approaches are fully
phenomenological. In particular the minimum of those potentials is
systematically occurring below saturation density in the Skyrme
functionals \cite{Raduta2014} for the considered stiffness coefficients, while in the RMF models we have
analyzed, it systematically occurs above saturation. 
In particular, in RMF models the scalar fields saturate at large
densities. The role of the scalar isoscalar meson, and scalar densities on
the properites of  RMF models has been discussed several times in the past, see
\cite{walecka,brito}.

Since the instability at high baryonic density seems to be strongly
correlated with the $NY$ and $YY$ potentials at low hyperonic
densities, there is hope that upon confronting the different functional forms 
to new more extensive experimental hypernuclear data to solve the
ambiguity. This perspective is left for future work.

\begin{figure*}
\begin{center}
\includegraphics[angle=0, width=0.48\textwidth]{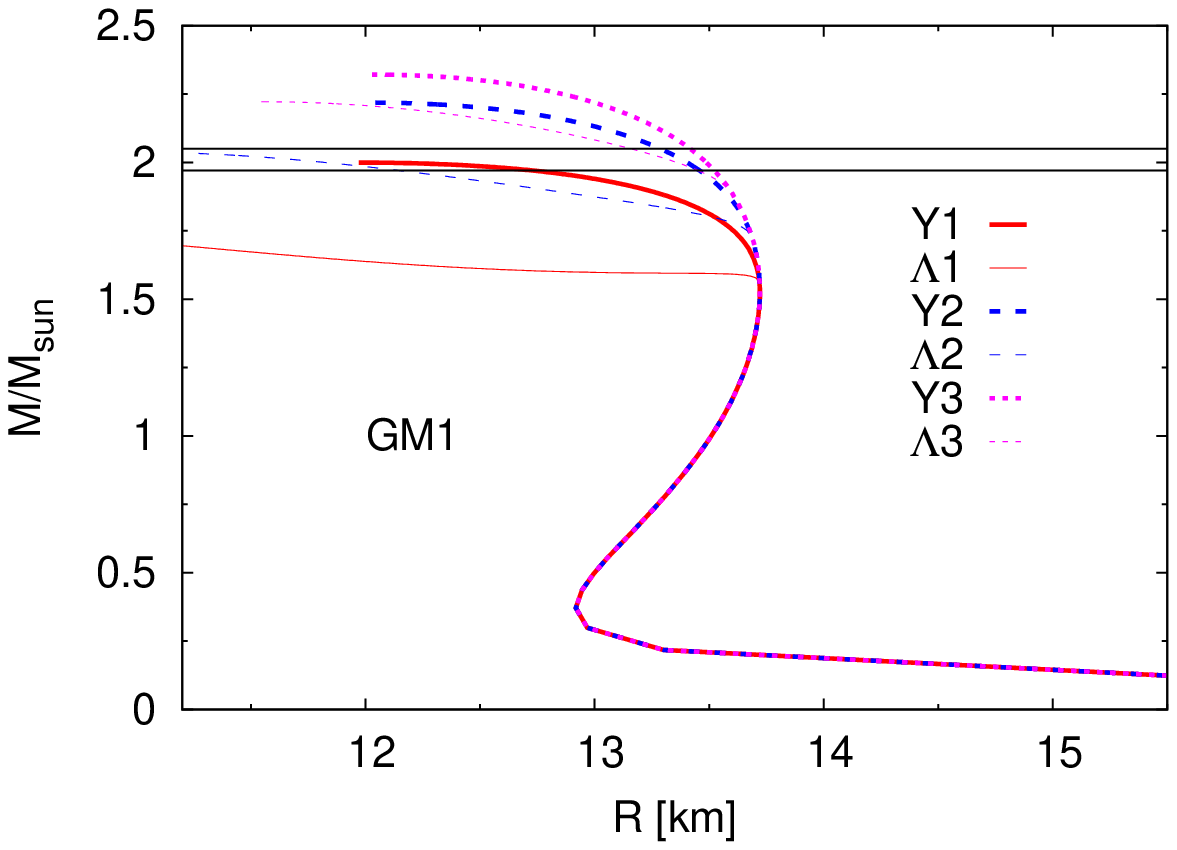}\hfill
\includegraphics[angle=0, width=0.48\textwidth]{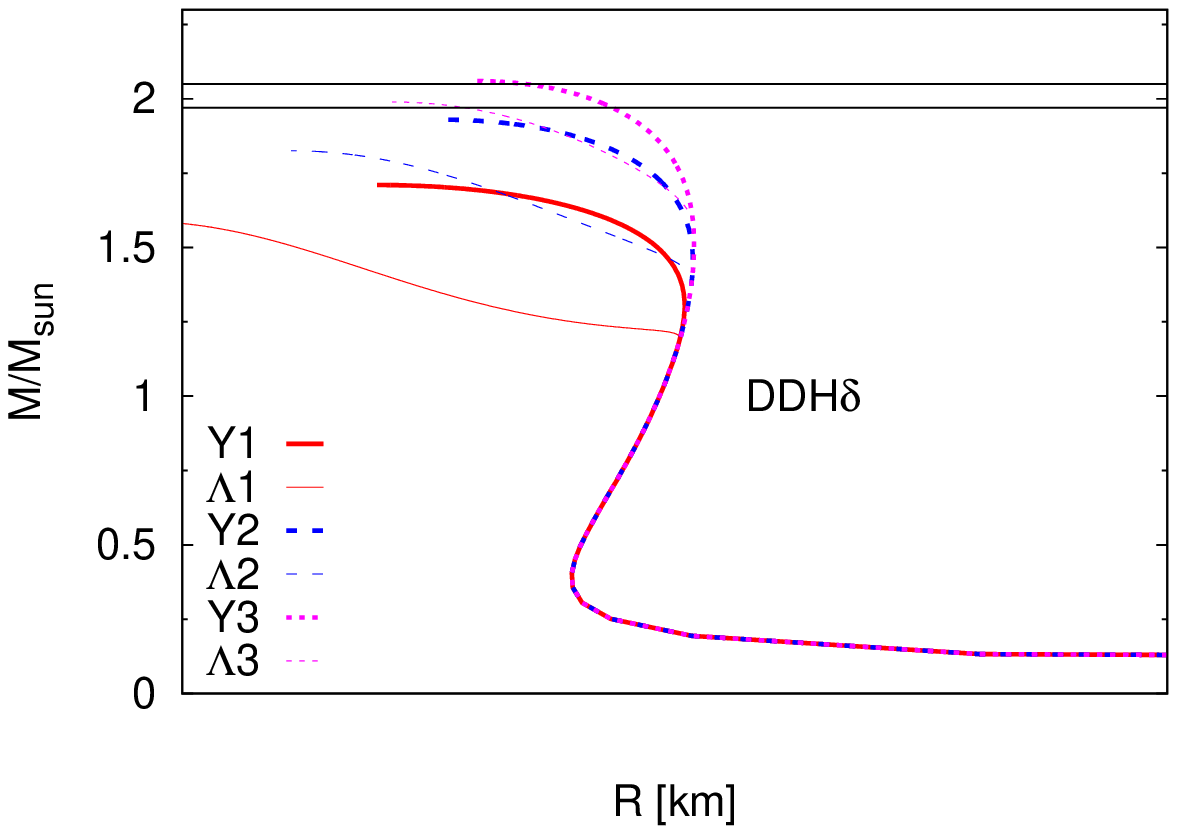}
\end{center}
\caption{\it Mass-radius relation for non-rotating spherical neutron
  stars. The GM1 (left) and DDH$\delta$ (right) parameter sets have been
  employed. The setup is the same as in Fig.~\ref{fig:lambdasymcritical}.The two horizontal lines
  indicate the mass of PSR J0348+0432, 2.01$\pm 0.04 M_\odot$.  }
\label{fig:lambdasymtov}
\end{figure*}


\begin{figure*}
\begin{center}
\includegraphics[angle=0, width=0.48\textwidth]{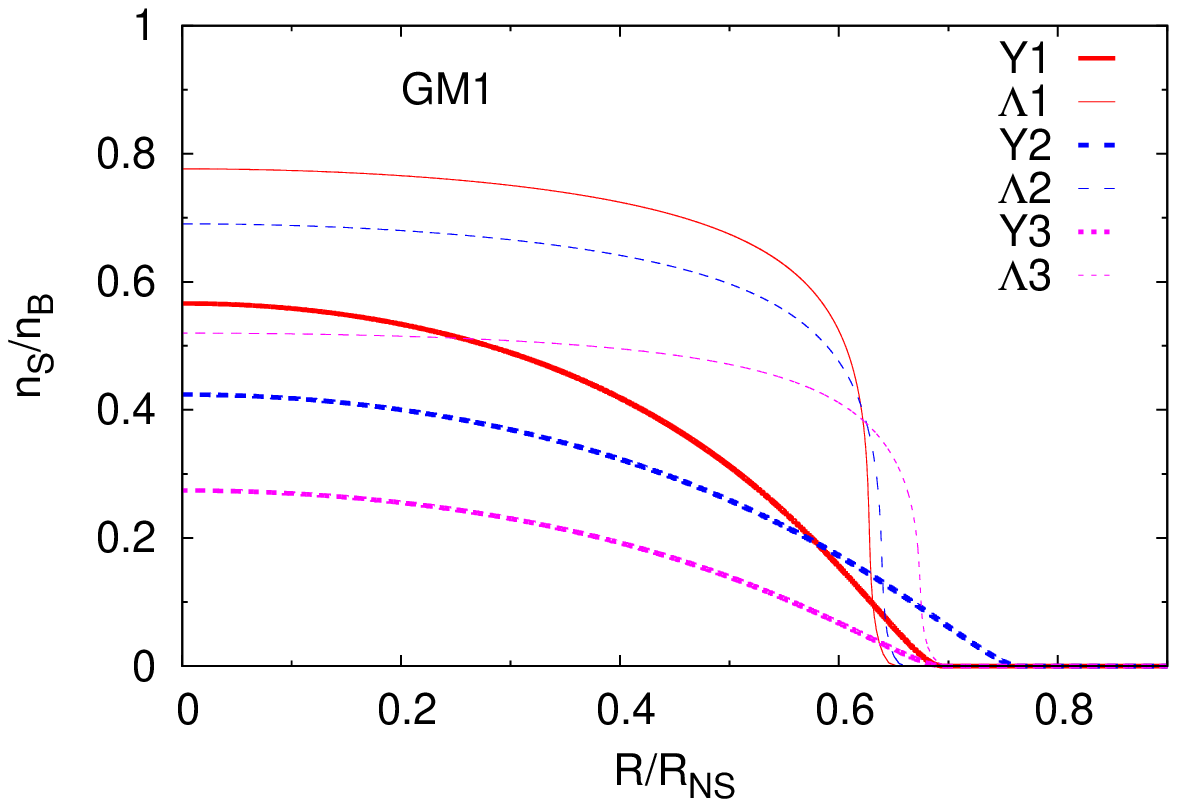}\hfill
\includegraphics[angle=0, width=0.48\textwidth]{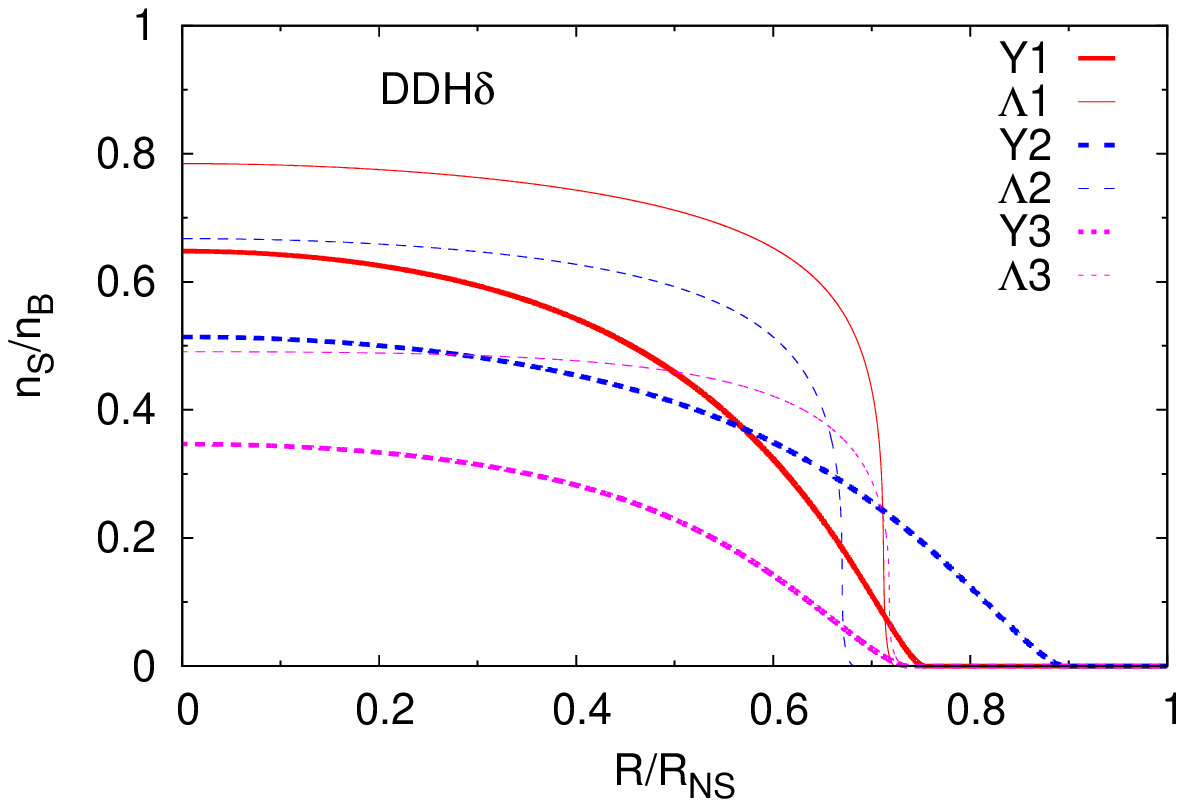}
\end{center}
\caption{\it The ratio of strangeness density ($= n_{\Lambda}$ for the present
  case) to baryon number density, $Y_s = n_S/n_B$ as a function of radius,
  normalized to the star's radius, for the respective maximum mass
  configuration of a spherical star. 
The GM1 (left) and DDH$\delta$ (right) parameter
  sets have been employed. The setup is the same as in
  Fig.~\ref{fig:lambdasymcritical}.}
\label{fig:lambdasymprofilefs}
\end{figure*}

\subsubsection{Neutron star masses and radii}
\label{sec:lambdaneutronstar}

In order to obtain neutron star masses and radii we solve the TOV
equations~\cite{TOV} for hydrostatic equilibrium of a non-rotating (spherical)
star in general relativity with the different equations of state (EoS). The
different EoS have been supplemented at low densities with the crust EoS from
Baym, Pethick and Sutherland~\cite{bps}. The results for the mass-radius
relation are displayed in Fig.~\ref{fig:lambdasymtov} for GM1 (left) and
DDH$\delta$ (right) following the symmetry inspired procedure. The two values
of $g_{\sigma^* \Lambda}$ for each value of $z$ thereby correspond to the
cases discussed above: the one leading to the canonical value of
$U_{\Lambda}^{(\Lambda)} (n_0/5) = -5 $MeV and the one corresponding to the
critical value for the onset of an instability. Qualitatively the results look
very similar in both models.

Concerning the maximum mass associated with the different EoS, it is obvious
that decreasing $z$, the maximum mass increases. This finding is not new, see
Ref.~\cite{Weissenborn2012}, and is explained by the fact that a smaller $z$
leads to an interaction with stronger repulsion at high densities. Similarly,
increasing $g_{\sigma^* \Lambda}$ renders the interaction more attractive and
lowers therefore the maximum mass. In both models, maximum masses compatible
with the recent observations of neutron stars with masses of $1.97 \pm
0.04$~\cite{Demorest10} and $2.01 \pm 0.04$~\cite{Antoniadis13} can be
obtained. The effect of a nonzero coupling to $\sigma^*$ is to reduce the
allowed parameter range in $z$ and smaller values of $z$ are required to
obtain a high enough maximum mass. The weak attraction suggested by
experimental data still allows for a wide range in $z$, whereas the strong
attraction leading to an instability reduces the allowed range in $z$
considerably. 

The latter conclusion can be softened relaxing the symmetry constrained on the
isoscalar vector couplings, see Sec.~\ref{sec:fulltov} where several examples are shown with accepatble neutron star masses and at the same time to the onset of an instability. 


It has been claimed that the strong repulsion needed within RMF models as well
in the hyperonic sector as for the purely nuclear part to obtain neutron stars
compatible with recent mass measurements and containing hyperons would lead
always to very large radii and that there would be a tension with recent
radius determinations, see for instance~\cite{Guillot13}. Let us stress at
this point that the radius determinations are difficult and that they are
presently far from being as reliable as the mass observations from
Refs.~\cite{Demorest10,Antoniadis13}.  The main problem is that the extraction
of radii from observations is much more model-dependent than the above
mentioned mass determinations, see e.g. Ref.~\cite{Heinke14}, where a
reanalysis gives a radius of $9.0^{+2.9}_{-4}$ km instead of
$6.6^{+1.2}_{1.1}$ km~\cite{Guillot13} for the same object, a neutron star in
NGC6397. A summary and discussion of different observational radius
determinations can be found in Ref.~\cite{Fortin2014}.  In addition, for a
rotating star due to its deformation there is no unambiguous relation between
the observed quantity and the radii determined theoretically. On the
theoretical side, due to the matching between a core and a crust EoS, not
necessarily obtained within the same model, the calculated radii are subject
to uncertainties of the order of several percent, too~\cite{Ishizuka2014,Providencia2014}.  However, much observational effort is put into neutron
star radii and further constraints are to be expected. Thus, it is interesting
to investigate the radius range neutron stars with hyperons can have.

The first remark to be made, looking on the radii in
Fig.~\ref{fig:lambdasymtov}, is that the central density exceeds the
threshold for the onset of $\Lambda$-hyperons only for neutrons stars
with $M \gsim 1.5 M_\odot$ (GM1) and $M \gsim 1.4 M_\odot$
(DDH$\delta$), respectively. Thus the radius at the canonical mass of
$M = 1.4 M_\odot$ is almost exclusively determined by the nuclear part
of the EoS, i.e. the nuclear interaction. The finding that the radius
(see table~\ref{tab:lambdasym} for a summary of the different values)
for $M= 1.4 M_\odot$ is significantly higher in GM1 than in in
DDH$\delta$ is consistent with it being dominated by the nuclear
interaction and shows again the strong impact of the incompressibility
and the symmetry energy $a_\mathit{sym}$ and its slope $L$, see
table~\ref{tab:nuclear}, which are considerably lower in DDH$\delta$
than in GM1. It is, however, not true that hyperons cannot be added to
nuclear models with low incompressibility or symmetry energy and slope
without violating the neutron star maximum mass constraint, see the
examples in DDH$\delta$ in Fig.~\ref{fig:lambdasymtov}, see
Table~\ref{tab:lambdasym} We will further discuss this point in
Section~\ref{sec:fulltov} upon including the full baryonic octet.

\subsubsection{Strangeness content of neutron star matter}
Does the stiffening of the EoS necessary to obtain maximum masses of at least
$\sim 2 M_\odot$  reduce the hyperon content of neutron star matter finally
excluding hyperons from neutron stars? The general recipe to increase the
maximum mass is clear: add additional short range repulsion. If this is done
mainly in the hyperon sector, then the strangeness content will be
reduced. This is what happens upon decreasing $z$. In
Fig.~\ref{fig:lambdasymprofilefs} we display the ratio of
strangeness density with respect to baryon number density in GM1 (left panel)
and DDH$\delta$ (right panel) as function of radius for the maximum mass
configurations obtained earlier. 

The general trend confirms the findings of Ref.~\cite{Weissenborn2012},
decreasing $z$ decreases the hyperon content of neutron star matter. There is,
however, a point to add. The hyperon content, as can be seen from
Fig.~\ref{fig:lambdasymprofilefs} is very sensitive to the attraction
furnished by a coupling to $\sigma^*$. Of course, adding attraction reduces
again the maximum mass such that the general trend is not modified: a higher
maximum mass means globally less hyperons. But the absolute value of the
hyperon content is strongly model dependent. And, as can be seen from the
examples in Fig.~\ref{fig:lambdasymprofilefs} present observations are far
from excluding hyperons from neutron stars.

\subsection{Including the full octet}
\label{sec:fulltov}
In the previous section, only $\Lambda$-hyperons have been
considered. Although $\Lambda$-hyperons are the first to appear in neutron
star matter for most interactions (in particular for models with a repulsive
$\Sigma^-$ potential in symmetric nuclear matter) and are in general the most
abundant hyperons, this is of course not completely realistic. Therefore, we
will now repeat the same analysis, but allowing a priori all particles of the
baryon octet to have a nonzero density. In the following subsections
we will first perform a stability analysis, and then  look at neutron
star properties.

\subsubsection{Stability analysis}

\begin{figure}
\includegraphics[ width=0.8\linewidth ]{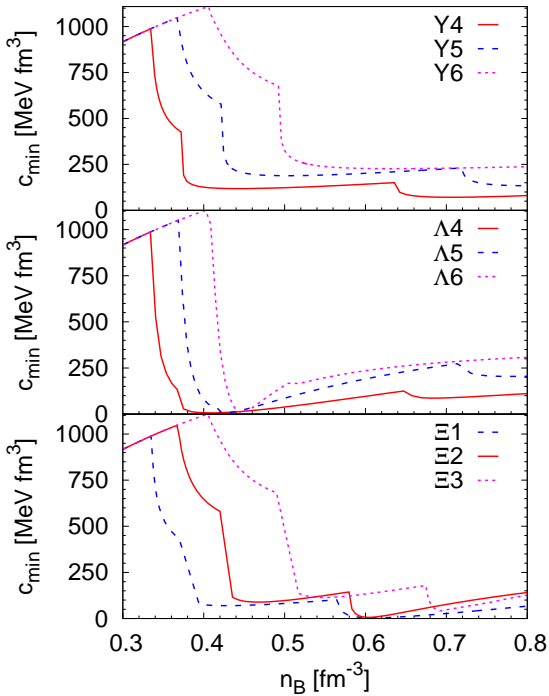}
\caption{\it Smallest eigenvalue of the curvature matrix of the energy density as
  a function of baryon number density for neutron star matter in  GM1 for different parameter sets including the full baryonic
  octet. Top: parameter sets not showing any instability, middle:
  instability driven by the onset of $\Lambda$, bottom: instability driven by
  the onset of $\Xi$. Parameters and neutron star properties are given in Table~\ref{tab:gm1allsym}.}
\label{fig:cminallsym}
\end{figure}

\begin{figure}
\includegraphics[width=0.8\linewidth,angle=0]{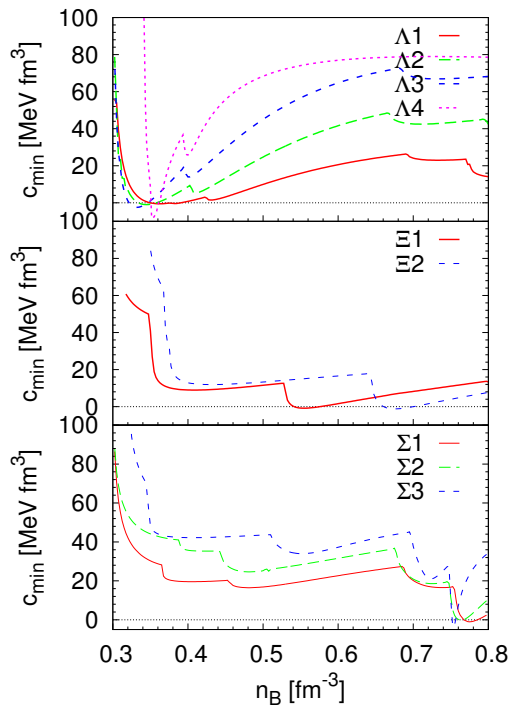}\\
\caption{\it Smallest eigenvalue of the curvature matrix of the energy density
  as a function of baryon number density for neutron star matter in TM1-2 for
  different parameter sets including the full baryonic octet. Top: parameter
  sets with an instability driven by the onset of $\Lambda$, middle:
  instability driven by the onset of $\Xi$, bottom: instability driven by the
  onset of $\Sigma$. Parameters and neutron star properties are given in Table~\ref{tab:tm1cassig}. }
\label{fig:cminalla}
\end{figure}

\begin{figure*}
\begin{tabular}{cc}
\centering
\includegraphics[width=0.5\linewidth,angle=0]{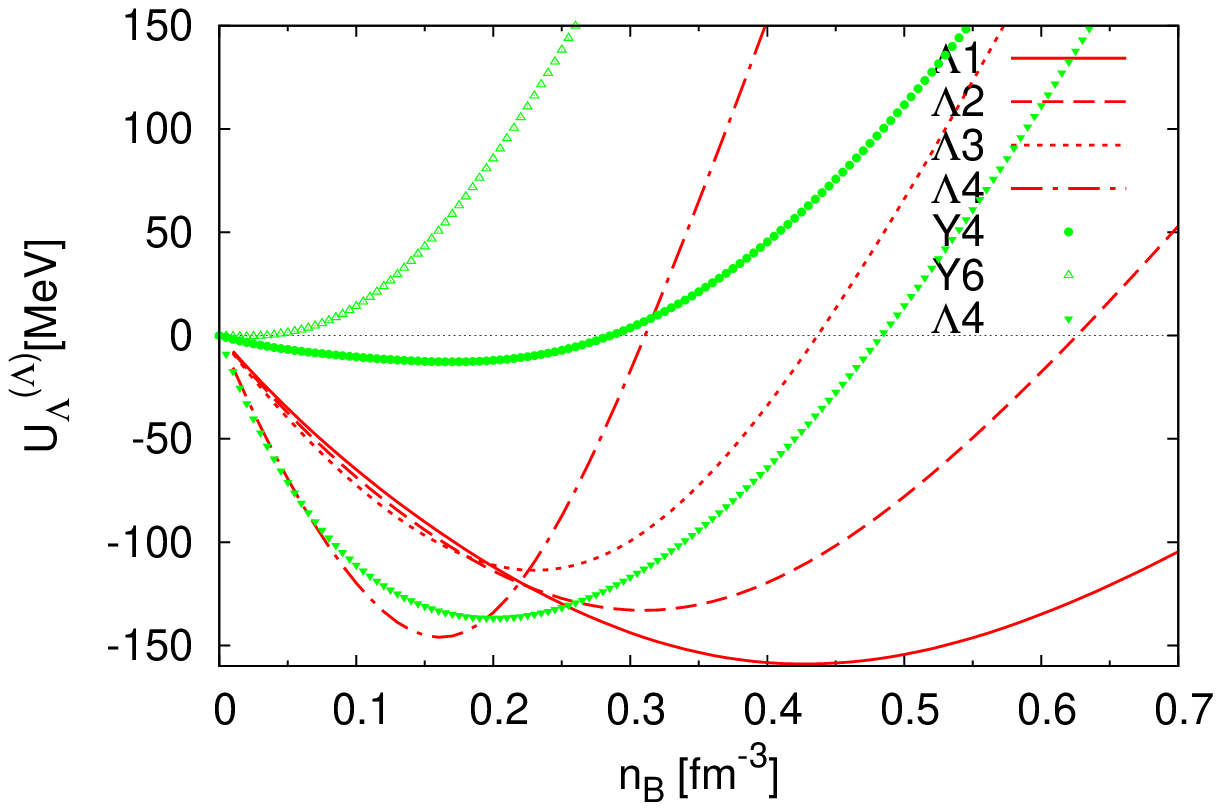}&
\includegraphics[width=0.5\linewidth,angle=0]{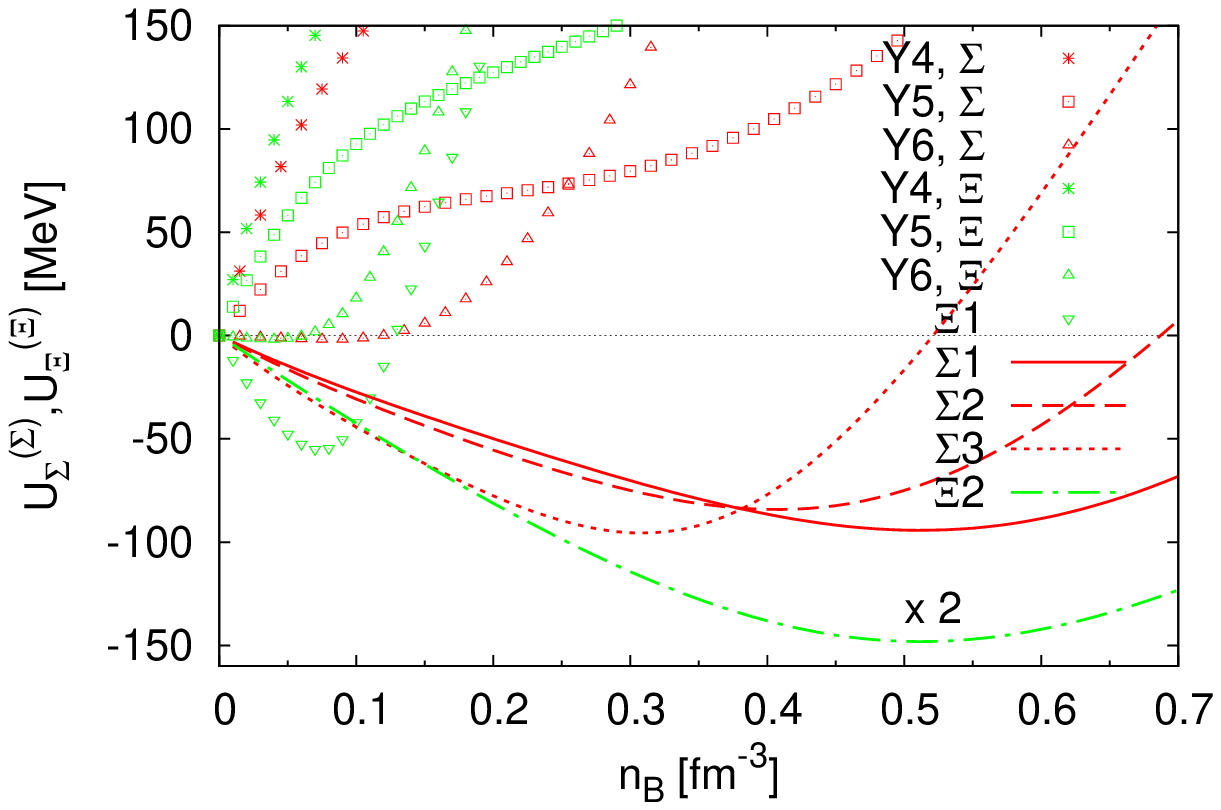}
\end{tabular}
\caption{\it The $Y$-potential in isospin symmetric $Y$-matter, for
  the $\Lambda$ (left) and $\Sigma$ and $\Xi$ (right) hyperons,  and for
  TM1-2 (lines)
  and
  DDH$\delta$ (symbols)  as a function of baryon
  number density for a selection of parameter sets in
  Table~\ref{tab:tm1cassig}. The results for model $\Xi 2$ should be
  considered multiplied by a factor of 2.}
\label{fig:upotentialsall}
\end{figure*}

\begin{table*}
\begin{tabular}{l|cccccccccccccc}
\hline
Model &$R_{\sigma^* \Lambda}$ & $R_{\sigma^* \Xi}$ &$R_{\sigma^* \Sigma}$
& 
$z$ & $M_{\mathit{max}}$ & $R_{1.4}$ & $R_{\mathit{max}}$
&$\epsilon^{(c)}$&$n_B^{(c)}$& $Y_S^{(c)}$&$f_S$ 
& $U_{\Lambda}^{(\Lambda)}(n_0/5) $  &
$U_{\Xi}^{(\Xi)}(n_0/5) $ & $U_\Sigma^{\Sigma} (n_0/5)$
\\
& &&&& $[M_\odot]$ & [km]& [km]&[MeV fm$^{-3}$]&[fm$^{-3}$]& &&[MeV]&[MeV]&[MeV] \\ \hline \hline
GM1\\
 $Y4$ &0.45 &1.16 & 0&0.41 &1.79  & 13.8 & 13.0   & 825.8  &0.73 &0.70&0.04&-5 &-10& 14\\
 $\Lambda 4$& 0.91&1.16& 0 & 0.41 & 1.59  & 13.8 & 13.6   & 598.3  & 0.56&0.59&0.006&-19
 &-10 & 14\\
 $\Xi 1$ & 0.45 & 1.35 & 0& 0.41 & 1.71  & 13.8  & 13.4   & 737.1  &0.67 &0.75&0.02&-5
 &-21 & 14\\ \hline
 $Y5$ &0.20& 0.87 & 0& 0.2 & 2.12  &13.8  & 12.3   & 1040  & 0.85& 0.71 & 0.07&-5&-10 &14\\
$\Lambda 5$ & 1.0& 0.87 & 0& 0.2 & 1.94  &13.8  & 11.0   & 491.1  &1.08 &0.93&0.20&-27 &-10 &14\\
$\Xi 2$ &0.20& 1.26 & 0& 0.2 &1.98   &13.8   & 13.1  &  806.9 & 0.71&0.84&0.03& -5&-29 &14 \\ \hline
 $Y6$ &0&0.55&0& 0. &2.29   &13.8  & 12.1   &1075   & 0.85&0.46&0.04& -7&-10 & 13\\
 $\Lambda 6$&1.08&0.55&0&  0. &2.19   & 13.8 & 11.7   & 1139  &0.90 &0.64&0.12& -34 &-10 &13 \\
 $\Xi 3$&0&1.13&0& 0. &2.24   &13.8   & 12.4   & 1004  &0.82  &0.76 &0.05&-7 &-33 &13 \\ \hline
\end{tabular}
\caption{Summary of results using the symmetry arguments in the
  isoscalar vector sector, meaning that the isovector couplings are fixed by
  the value of $z$ and the relation in Eqs.~\ref{eq:symmetry}. The
  parametrization GM1 has been employed. The parameters are indicated in
  column 2-5. 
 The values of $R_{\sigma^*\Lambda}$ and $R_{\sigma^*
    \Xi}$ for models $\Lambda n$ and $\Xi n$ thereby correspond to the
  critical values of these coupling constants for
  the onset of an instability, in the $\Lambda$ and $\Xi$-channel,
  respectively.
The central energy density, baryon number density and strangeness
  fraction $Y_s = n_S/n_B$ are given for the maximum mass configuration. $f_S$
  represents the integral of the strangeness fraction $Y_s/3$ over the whole
  star for the maximum mass configuration as in Ref.~\cite{Weissenborn2012}. 
   \label{tab:gm1allsym}  }
\end{table*}

As in Sec.~\ref{sec:lambdastable} we are interested here in sets of couplings
describing stellar matter with an instability at the onset of hyperons.  For a
given choice of the isoscalar vector meson couplings $g_{\phi i}$ and/or
$g_{\omega i}$ and fixing the couplings to $\sigma$ by the values of the
hyperon potentials in nuclear matter, the only remaining parameters are the
couplings to $\sigma^*$. The isovector vector couplings are kept fixed by
isospin symmetry. 

In Table~\ref{tab:gm1allsym} for
GM1 and Table~\ref{tab:tm1cassig} for TM1-2 and DDH$\delta$ we adopt
the following convention: if no instability is present the model is
identified with $Yn$; the sets identified with $\Lambda n$, all lead to an instability at the
onset of $\Lambda$;   the sets identified with $\Xi n$ and $\Sigma n$ originate
an instability driven by the onset of the $\Xi$ or the $\Sigma$ isovector
multiplet.

Let us start with the results respecting the symmetry constraints of
Eqs.~\ref{eq:symmetry}. We will show only results obtained using GM1 here for
two reasons. First, as discussed in Sec.~\ref{sec:lambda} qualitatively the
results are similar for different models. Secondly, as shown in
Sec.~\ref{sec:lambdaneutronstar}, for the models including only the
$\Lambda$-hyperon, the parameter space for obtaining high enough neutron star
masses is very reduced within DDH$\delta$ upon applying the symmetry
constraints to the isoscalar vector couplings. The allowed parameter space
becomes even smaller if the full octet is considered and the results do not
give us any new insight with respect to those with GM1 presented here. 

In Fig.~\ref{fig:cminallsym} we show the minimal eigenvalue of the curvature
matrix, $c_{\mathit{min}}$, as function of baryon number density in neutron
star matter for different choices of the couplings to $\sigma^*$ and $z$
within GM1.  The choice of couplings in the upper panel corresponds to those
giving $U_{\Lambda}^{(\Lambda)} (n_0/5) = -5$ MeV and $U_{\Xi}^{(\Xi)}(n_0/5)
= -10$ MeV.  Since only $\Lambda, \Xi^-$, and $\Xi^0$-hyperons appear in
neutron star matter with the employed parameter set, only the couplings
associated to these hyperons are given. The $\Sigma$-hyperons, mainly due to
the assumed strongly repulsive $\Sigma N$ interaction, appear only well above
$n_B=$ 1 fm$^{-3}$ beyond the central density of the neutron stars with the
highest mass. In the curves successive thresholds, leading to kinks in
$c_{\mathit{min}}$, can be observed. They correspond to the onset of $\Lambda,
\Xi^-$, and $\Xi^0$-hyperons, respectively.

For the above choice of parameters, the system is perfectly stable. However, we
have seen before for the case of nuclear matter with $\Lambda$-hyperons, see
Fig.~\ref{fig:lambdasymcritical}, that increasing $g_{\sigma^* i}$ decreases
the minimal eigenvalue of the curvature matrix leading finally to an
instability.  This can again be observed here. In the middle and bottom
panels, results are displayed with the smallest value of $g_{\sigma^*
  \Lambda}$ and $g_{\sigma^* \Xi}$, respectively, leading to an instability
for a given choice of the other couplings. As can be seen in the bottom panel,
within GM1, for the critical value of $g_{\sigma^*\Xi}$, the system is not
driven into an instability when the $\Xi^-$ sets in, but the instability
arises rather at the $\Xi^0$-threshold. In addition, although the $\Lambda$ is
the first hyperon to appear, after the onset of $\Xi^-$, the number of
$\Lambda$ hyperons remains almost constant with increasing density and starts
even to decrease with the onset of
$\Xi^0$ due to the large attraction that $\Xi^-$ and $\Xi^0$ feel induced by
the large coupling to $\sigma^*$.  Changing the isovector channel by choosing
a smaller symmetry energy slope would move the instability to larger
densities, because a smaller $L$ disfavors the onset of neutral hyperons
\cite{providencia13}. We will discuss this statement in a more quantitative
way within two versions of TM1-2 below. 

The critical values for $g_{\sigma^* \Lambda}$ leading to an instability are
slightly lower than those obtained in Sec.~\ref{sec:lambdastable} for
$n,p,\Lambda,e$-matter, except for the $z=0$ case. On a first sight this might
be surprising since at densities below the $\Xi^-$-threshold, the results
should be exactly the same. The reason is that actually the instability is
here not caused by the onset of $\Lambda$-hyperons, but $\Xi^-$-hyperons. The
threshold densities are very close and the two distinct thresholds are hardly
distinguishable looking at $c_{\mathit{min}}$. A closer inspection of the data
shows that the minimal value of $c_{\mathit{min}}$ lies at densities above the
$\Xi^-$-threshold, see Fig.~\ref{fig:gm1fractions}, too, where the number
fractions are shown for the different species.

Not astonishingly, we are able to find an instability in other models, too. In
Fig.~\ref{fig:cminalla} several examples of parameter sets leading to an
instability within TM1-2 and DDH$\delta$ are shown. The $g_{\sigma Y}$
couplings are adjusted to the hyperon potentials in nuclear matter as
before. Again, the values of $g_{\sigma^* i}$ correspond to the limiting
values for the onset of an instability, as seen from the behavior of
$c_{\mathit{min}}$.  No symmetry constraints have been imposed on the
isoscalar vector couplings. All parameter values are listed in
Table~\ref{tab:tm1cassig}.

The sets in Table~\ref{tab:tm1cassig} identified with $\Lambda n$,
$\Xi n$ or $\Sigma n$,
displayed in Fig.~\ref{fig:cminalla} on the top, middle and bottom
panels, respectively, all lead to an instability at the
onset of $\Lambda$, $\Xi$ or $\Sigma$  hyperons. 
 A $\Xi$-driven instability is possible for strong $g_{\sigma^* \Xi}$
and weak $g_{\sigma^* \Lambda}$ couplings. A $\Sigma$-driven instability
requires a strong $g_{\sigma^* \Sigma}$ coupling.  Just as with GM1, a $\Xi$
or $\Sigma$ driven instability was only obtained after all members of the
multiplet set in. It should be pointed out that, although the $\Sigma$ driven
instability occurs at a quite high density, it still occurs within the range
of densities inside a neutron star, see Secs.~\ref{sec:allneutronstar} and
\ref{sec:strangenessall}. A less repulsive $U_\Sigma^{(N)}$ would allow for an
instability at lower densities. In the same way a less attractive
$U_\Xi^{(N)}$ would remove the onset of the $\Xi$ inside a neutron star.

Let us mention that, in contrast to GM1, neutron star matter contains
$\Sigma$-hyperons for some sets in Table~\ref{tab:tm1cassig} with
$U_\Sigma^{(N)} = 30$ MeV and more than three thresholds can be observed
within the range of densities relevant for neutron stars. We will discuss the
composition in detail in Sec.~\ref{sec:strangenessall}.

Of course, the $YY$-interaction is very sensitive to the couplings to
$\sigma^*$. Remember that originally the $\sigma^*$ has been
introduced to allow for very attractive $YY$-interactions in view of
experimental results for double-$\Lambda$ hypernuclei at the epoch in
Ref.~\cite{Schaffner1994}. A strong attraction in the
$\Lambda\Lambda$-channel was indeed found in
Sec.~\ref{sec:lambdastable} for the values of $g_{\sigma^*\Lambda}$
leading to an instability. The same is true here, as seen for the
$\Xi\Xi$-potentials and the $\Sigma\Sigma$-potentials, shown in
Fig.~\ref{fig:upotentialsall}, right hand side, and the
$\Lambda\Lambda$-potential displayed in Fig.~\ref{fig:upotentialsall},
left panel, for different examples in TM1-2 and DDH$\delta$. The
corresponding values at $n_0/5$ are listed in
Table~\ref{tab:tm1cassig}. It is obvious that the attraction needed in
the $\Lambda\Lambda$-channel to obtain an instability is much higher
for all examples shown than the values suggested by experimental
data. Even if we neglect the Nagara event and only consider the
earlier data (see e.g.~\cite{Schaffner1994} for a discussion)
resulting in a stronger attraction in this channel, the coupling
strength needed for the onset of an instability is far off.

In the $\Xi\Xi$-channel and the $\Sigma\Sigma$-channel the situation
is less evident because there is no experimental information available
in these channels.  Current information, based on theoretical
arguments for the baryon octet in vacuum and corresponding meson
exchange models, is clearly not sufficient to pin down the amount of
attraction for the $YY$ interaction in dense matter.  Since the
coupling to $\sigma^*$ is determined mainly via the $YY$-interaction,
more data, in particular, on other hyperons than $\Lambda$-hyperons
would be very welcome to be able to judge whether the different chosen
values are pertinent or not.

\begin{table*}
\begin{tabular}{l|lcccccccccccccc}
\hline
 Model&$L$ & $R_{\sigma^* \Lambda}$&$ R_{\sigma^* \Xi}$ &$R_{\sigma^* \Sigma}$& 
$R_{\omega Y}$ 
&$R_{\phi Y}$ 
& $M_{\mathit{max}}$ & $R_{1.4}$ & $R_{\mathit{max}}$ &$\epsilon^{(c)}$&  $n_B^{(c)}$
&$f_s$&  $U_{\Lambda}^{(\Lambda)}(n_0/5) $  &
$U_{\Xi}^{(\Xi)}(n_0/5) $ &
$U_{\Sigma}^{(\Sigma)}(n_0/5) $ 
\\
&[MeV]& & &&&&$[M_\odot]$ & [km]& [km]&[MeV fm$^{-3}$]&[fm$^{-3}$] & & [MeV]&[MeV]
&[MeV] \\ \hline \hline
TM1-2 \\
Y1&110&
0&0&0& 1& 1& 1.95 &14.55 &12.57& 1028&0.86 &0.15& 1.7&21.1 &16.17\\
Y2&$55$&
0&0&0&  1& 1& 1.94 &13.43 & 12.02& 1085&0.91 &0.12& 1.7&21.1 & 16.17\\ \hline
$\Lambda$1&110&1.01&1&1& 1&1 & 1.51 &14.55  &14.48& 505.2&0.48
&0.004&-21.8&-2.3 &-7.17\\
$\Lambda$2&110&1.23&1.23&1.23& 1& 1.5&1.74&14.55 &10.61&1634 &1.28&0.40&-23.8&23.1 &-0.84\\
$\Lambda$3&110&1.48&1.48& 1.48& 1 & 2 &1.90&14.55 & 11.04&1458 &1.15&0.31&-27.0&59.1&-12.62\\
$\Lambda$4&110& 1.68&1.68& 1.68& 1.5
& 2& 2.13&14.55 & 12.20&1119 &0.90& 0.16&-41.1&36.0 &-20.96\\ 
$\Lambda$5&55&1.445&1.445& 1.445& 1 &2&1.85&13.43&10.76&1482 &1.17&0.26
& -23.9&62.1 & -9.40\\
$\Lambda$6&55&1.58&1.58& 1.58&1.5 & 2&
2.09& 13.43&11.85&1337 &0.92&0.11&-33.7&44.4 &-12.77\\ \hline
$\Xi$1&$110$  & 0.3& 1.42& 1 &1 & 1 &1.75&14.63& 14.08&722.2&0.65 &
0.04&-0.37 & -25.82 & -7.17\\
$\Xi$2&$55$  & 0.3& 1.42& 1&1 &1 &1.78&13.43& 12.89&921.5&0.78 &
0.06&-0.37 & -25.26 &-6.93\\ \hline
$\Sigma$1&$110$& 0.8& 1&1.1 & 1 &1.2  &1.75& 14.57
&13.21&856.4&0.76&0.11&-9.80&11.04 & -9.68\\
$\Sigma$2&110& 0.8& 1&1.23 & 1 &1.5  &1.87&14.57&13.12&921.5&0.79&
0.10&-3.74&35.11 &-12.75\\
$\Sigma$3&110& 0.8& 1&1.37& 1.3 &1.5&
2.05&14.57&12.91&959.0&0.80&0.12&-4.19 &30.36& -14.66\\ 
\hline \hline
DDH$\delta$ \\
 Y4  & 44&1.03&0&0&1.5 &0.85 & 2.05 &12.7&11.2&1217& 0.99&0.04 & -5&79 &62\\ 
  Y5 &44&1.03 &1&1&1.5& 0.85& 2.00 &12.7& 11.1 &1262&1.02& 0.06&-5&41 &24 \\ 
  Y6 &44&1.81 &2.70&1.99&1.5& 1.59& 2.01 &12.7& 10.9 &1325&1.06& 0.08&-0.2&-1 &-0.4 \\ 
  \hline
  $\Lambda 4$ &44&1.5 &0&0&1.5& 0.85& 2.04 &12.7& 10.9 &1358&1.07& 0.08&
  -48&79 &62 \\ 
  $\Xi 1$ &44&1.81 &2.85&1.99&1.5& 1.59& 2.00 &12.7& 10.9 &1325&1.06& 0.09&
  -0.2&-33 &-0.4 \\ 

\hline

\end{tabular}
\caption{
  Summary of results calculated within different models allowing for a free
  variation of the isoscalar vector couplings. The parameters used
  are indicated in column 2-7. $R_{\omega \Lambda}$ and $R_{\phi \Lambda}$ represent the ratio
  of the corresponding isoscalar vector coupling constants to their respective
  $SU(6)$-values. As before, the couplings to $\sigma^*$ are defined with respect to $g_{\sigma N}$. The choices of
  parameters named $\Lambda n, \Xi n, \Sigma n$ with $g_{\sigma^*Y}\ne 0$ originate an instability driven by the onset of
  $\Lambda$,  $\Xi$ or $\Sigma$, respectively. The parametrizations named Y$n$
  do not show any instability. The central energy density and baryon number density are given for the maximum mass configuration. $f_S$
  represents the integral of the strangeness fraction $Y_s/3$ over the whole
  star for the maximum mass configuration. }
\label{tab:tm1cassig}
\end{table*}

\subsubsection{Neutron star masses and radii}
\label{sec:allneutronstar}
\begin{figure*}
\begin{tabular}{cc}
\includegraphics[width=0.5\linewidth,angle=0]{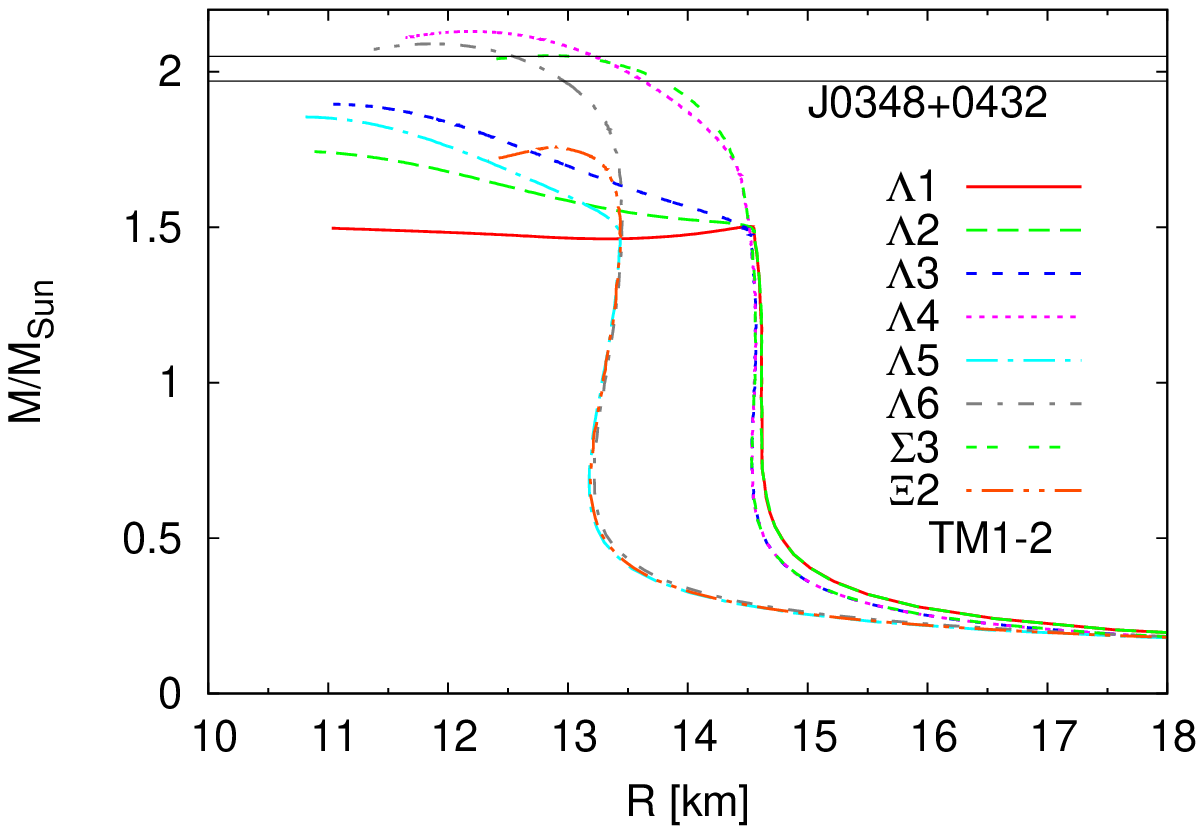}\hspace{0.cm}&
\includegraphics[width=0.5\linewidth,angle=0]{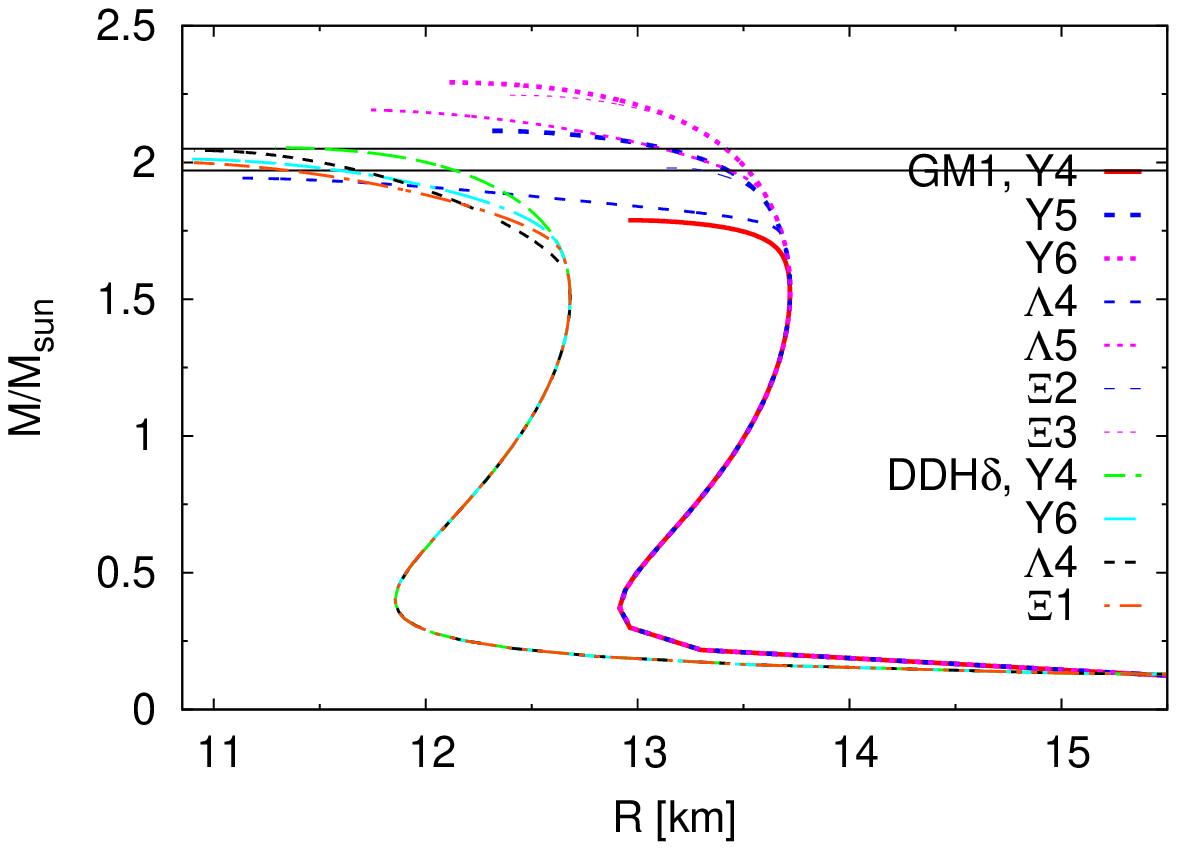}\\
\end{tabular}
\caption{\it (Gravitational) Mass/radius curves for spherical neutron stars
  for a selected choice of parameters using TM1-2
  (left) and GM1/DDH$\delta$ (right). For TM1-2, we show the families of stars
  obtained with the parametrizations of Table
  \ref{tab:tm1cassig} that drive the stellar matter into an
  instability upon the  onset of $\Lambda$. 
The two horizontal lines
  indicate the mass of PSR J0348+0432, 2.01$\pm 0.04 M_\odot$.  }

\label{fig:neutronstarsall}
\end{figure*}
In Sec.~\ref{sec:lambdaneutronstar} we have presented results for neutron star
masses and radii with matter containing neutrons, protons, $\Lambda$-hyperons
and electrons. The conclusions were that, within this restricted setup,
firstly the observed neutron star masses can  not be used to exclude
the existence of a first order phase transition to hyperonic matter in RMF
models. Secondly, the radii for intermediate mass neutron stars are most
sensitive to the properties of the nuclear interaction and, in contrast to
previous claims, it is possible to obtain masses in agreement with recent
observations for models containing a substantial amount of $\Lambda$-hyperons,
using a nuclear interaction with low $L$ leading to relatively small radii of
the order 12-13 km at intermediate masses.

Within this section we would like to investigate whether these conclusions
remain true about including the full baryonic octet. From the simple argument
that new degrees of freedom soften the equation of state we would expect the
mass constraint to become more difficult to fulfill, at least if more
than one
hyperon species becomes populated. This argument is, however, only strictly
valid for free Fermi gases without interaction and we thus have to study the
questions within different interaction models. 

In Tables~\ref{tab:gm1allsym} (GM1) and \ref{tab:tm1cassig} (TM1-2 and
DDH$\delta$) the maximum mass, the radius at a gravitational mass of $1.4
M_\odot$ and the radius at maximum mass for spherical non-rotating neutron
stars are given for the different parameter sets discussed in the preceding
section. In Fig.~\ref{fig:neutronstarsall} the corresponding mass-radius
relation are plotted, for TM1-2 on the left hand side and GM1 and DDH$\delta$
on the right hand side.

Very generally, increasing the magnitude of the vector-meson couplings allows
for larger masses due to the stronger repulsion. This can be observed for all
cases. Within GM1, respecting the symmetries, the maximum mass increases with
decreasing $z$, see Table~\ref{tab:gm1allsym}. For TM1-2 and DDH$\delta$ no
constraint is set on the isoscalar vector couplings and the maximum neutron
star mass increases upon increasing those values, see
Table~\ref{tab:tm1cassig}.  

Increasing the $\sigma^*$-couplings, leading to a stronger attraction, the
maximum mass decreases. This can again be observed within all the models
discussed here.  For instance, taking for the isoscalar vector couplings their
respective $SU(6)$ values, and choosing $g_{\sigma^*i}=0$, a maximum mass of
$\sim 1.95 M_\odot$ is obtained for both $L=110$ and 55 MeV within
TM1-2. Increasing $g_{\sigma^*i}$ reduces the maximum mass as expected, and
for $g_{\sigma^*\Lambda}^{crit}=1.01 \,g_{\sigma N}$ the maximum mass is 1.51
$M_\odot$. If, however, no constraint is set on the isoscalar vector
couplings, it is possible to choose a set of couplings which predict larger
maximum masses, also above 2 $M_\odot$, and still give rise to an instability
driven by the onset of strangeness.

Compared with
the case of $n,p,\Lambda$+ $e$-matter, see Sec.~\ref{sec:lambdaneutronstar}, the
maximum masses are reduced by the presence of other hyperon species, as
expected. The effect is more pronounced if more species enter and if their
respective threshold densities are considerably below the central density of
the maximum mass configuration. For instance, as can be observed from the GM1
results, for the highest value of $z$ the maximum mass is strongly reduced
with respect to the $n,p\Lambda$+$e$ case and only very small for $z=0$. The
reason is that for the respective maximum mass configurations,
$\Xi^0$-hyperons enter for $z=0.41$ and $z=0.2$ in addition to $\Lambda$ and
$\Xi^-$, whereas they are absent for the $z=0$ models, see
Fig.~\ref{fig:gm1fractions}, too. Another point should be mentioned concerning
these results: the effect of increasing the attraction due to a
$\sigma^*$-coupling between the canonical and the critical value is more
pronounced for the threshold to $\Lambda$-hyperons than for the $\Xi$. The
reason is that the difference between the canonical and the critical value in
the $\Xi$-channel is smaller than for the $\Lambda$-hyperons, partly because
at the $\Xi$ thresholds other hyperons are already present pushing the
instability.

From all the above discussed examples, it is clear that the existence of an
instability is not excluded by the neutron star maximum masses. The maximum
masses are more strongly dependent on the vector couplings than on the
$\sigma^*$- couplings for values between zero and the critical values, such
that the allowed parameter space is still large. In addition, as discussed in
Sec.~\ref{sec:lambdaneutronstar}, there is no evidence that nuclear EoS with
large $L$ have to be chosen to obtain maximum masses above $2 M_\odot$ with a
considerable amount of hyperons in the central part.

Since for almost all the parameter sets considered here, $\Lambda$-hyperons
appear first, the neutron star radii at intermediate masses still depend
mainly on the properties of the purely nuclear EoS and considerable
differences due to the presence of hyperons can be observed only for masses
close to the respective maximum mass. For the EoS giving acceptable
maximum masses, the difference in radii due to hyperons becomes clearly
visible for stars with masses above roughly 1.8 $M_\odot$. Therefore again, the
main parameter determining the radii at intermediate masses is the slope of
the symmetry energy, $L$, as found earlier in the context of purely nuclear
models~\cite{hor01,hor03,rafael11}. 

To illustrate this point, let us first compare the two versions of TM1-2. It
should again be noted that within those models hyperons are present only in
stars with $M\gtrsim 1.5 \, M_\odot$, such that the radius at $M = 1.4
M_\odot$ is determined by the nuclear parameters.  The original
parametrization TM1-2 has a very large value of the symmetry energy slope at
saturation $L$, and, not surprisingly the radius of a 1.4 $M_\odot$ star is
above 14 km. However, including a non-linear $\omega\rho$ term in the
Lagrangian density it is possible to decrease $L$. For $L=55$ MeV, a radius of
13.4 km is obtained. The DDH$\delta$ model has an even lower value of $L = 44$
MeV and, as discussed already in Sec.~\ref{sec:lambdaneutronstar}, the radius
of a star with $M = 1.4 M_\odot$ is 12.7 km. Hyperons can be added to the EoS
within this model without any contradiction to present neutron star mass
observations.

Hence, since the maximum masses are only
moderately influenced by the presence of other hyperons than $\Lambda$'s, we
conclude as before that hyperons can be added to nuclear models with low
symmetry energy and slope without violating the neutron star maximum mass
constraint, and that in this way radii between 12-13 km can be obtained for
neutron stars with the canonical mass of $1.4 M_\odot$, see the numbers given
in Table~\ref{tab:tm1cassig}.

\subsubsection{Strangeness content of neutron star matter}
\label{sec:strangenessall}
\begin{figure*}
\begin{tabular}{ccc}
\centering
\includegraphics[width=0.3\linewidth,angle=0]{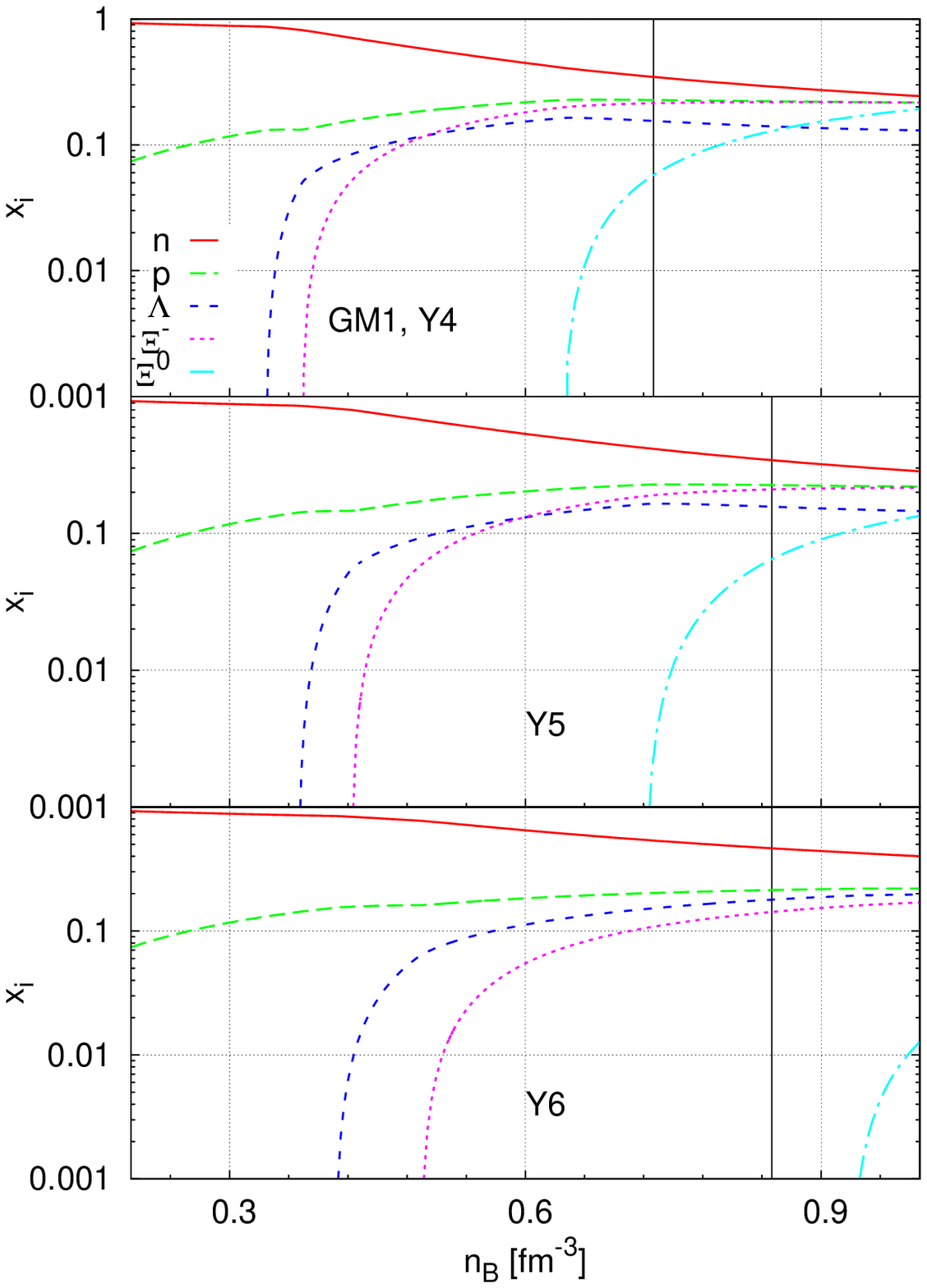}&
\includegraphics[width=0.3\linewidth,angle=0]{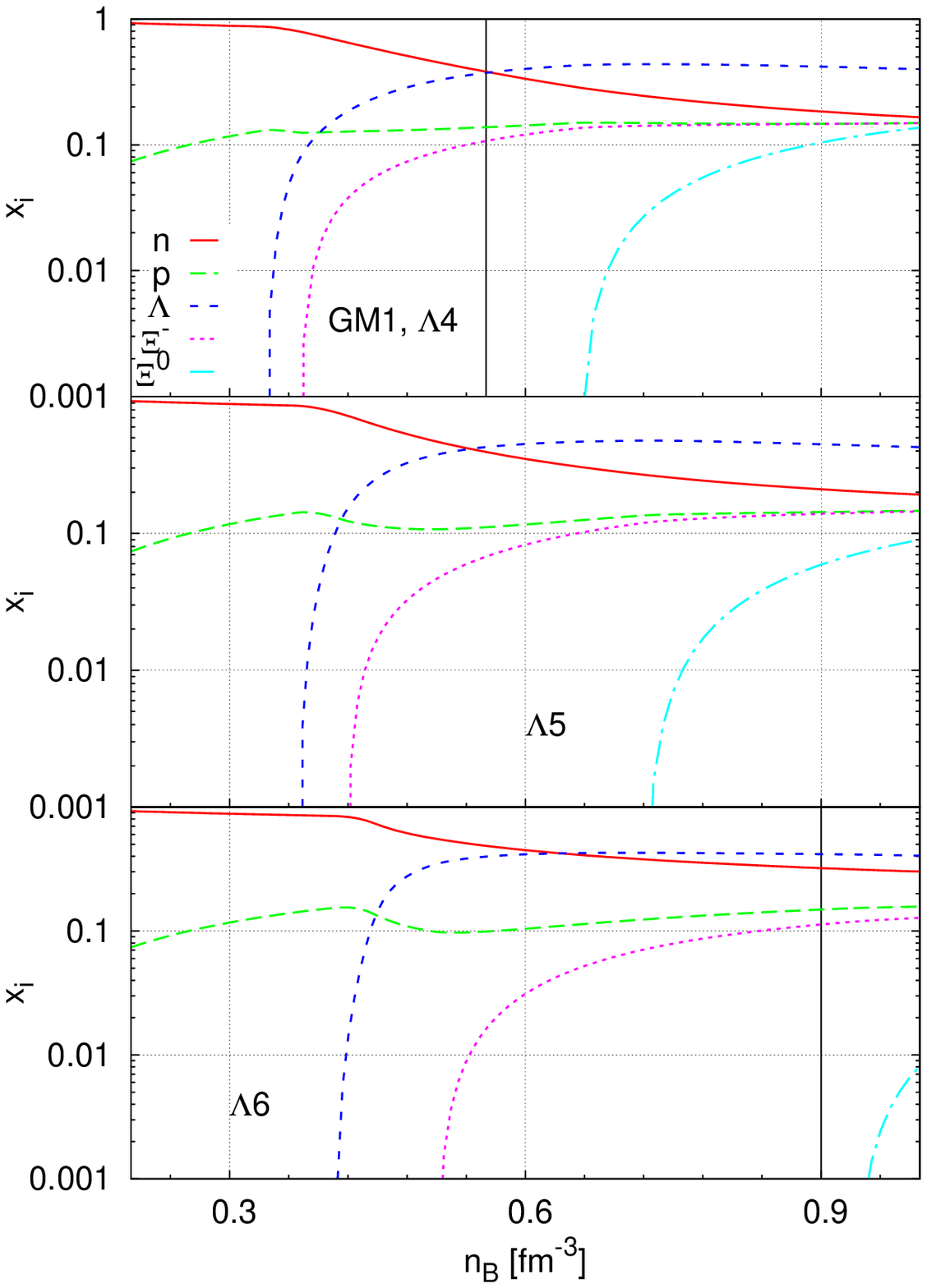}&
\includegraphics[width=0.3\linewidth,angle=0]{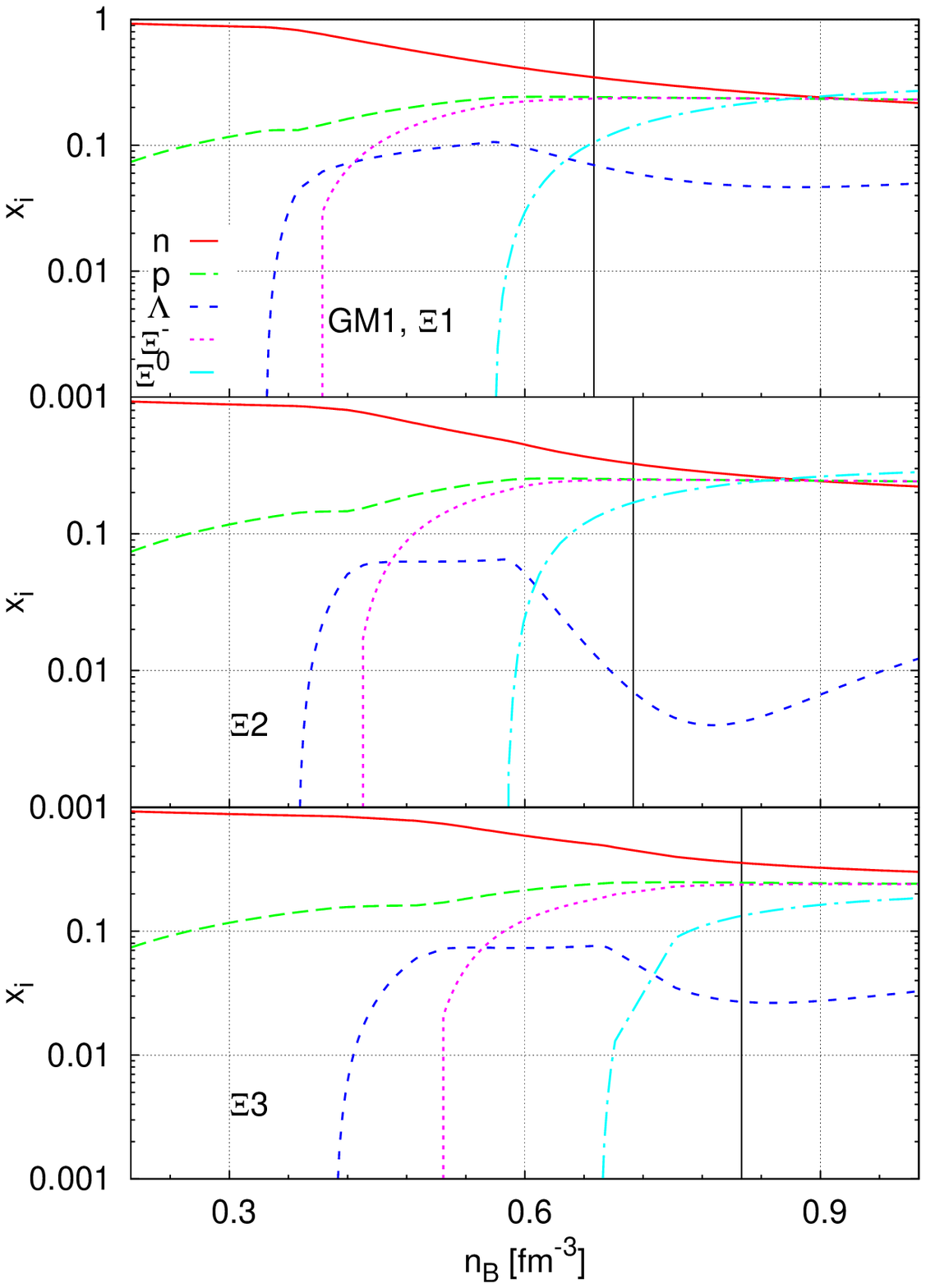}\\
\end{tabular}
\caption{\it Particle fractions in neutron star matter ($x_i = n_i/n_B$) for the
  different parameter sets discussed above within GM1. The vertical lines
  indicate the central density of the respective maximum mass configurations.}
\label{fig:gm1fractions}
\end{figure*}
\begin{figure*}
\begin{tabular}{ccc}
\centering
\includegraphics[width=0.3\linewidth,angle=0]{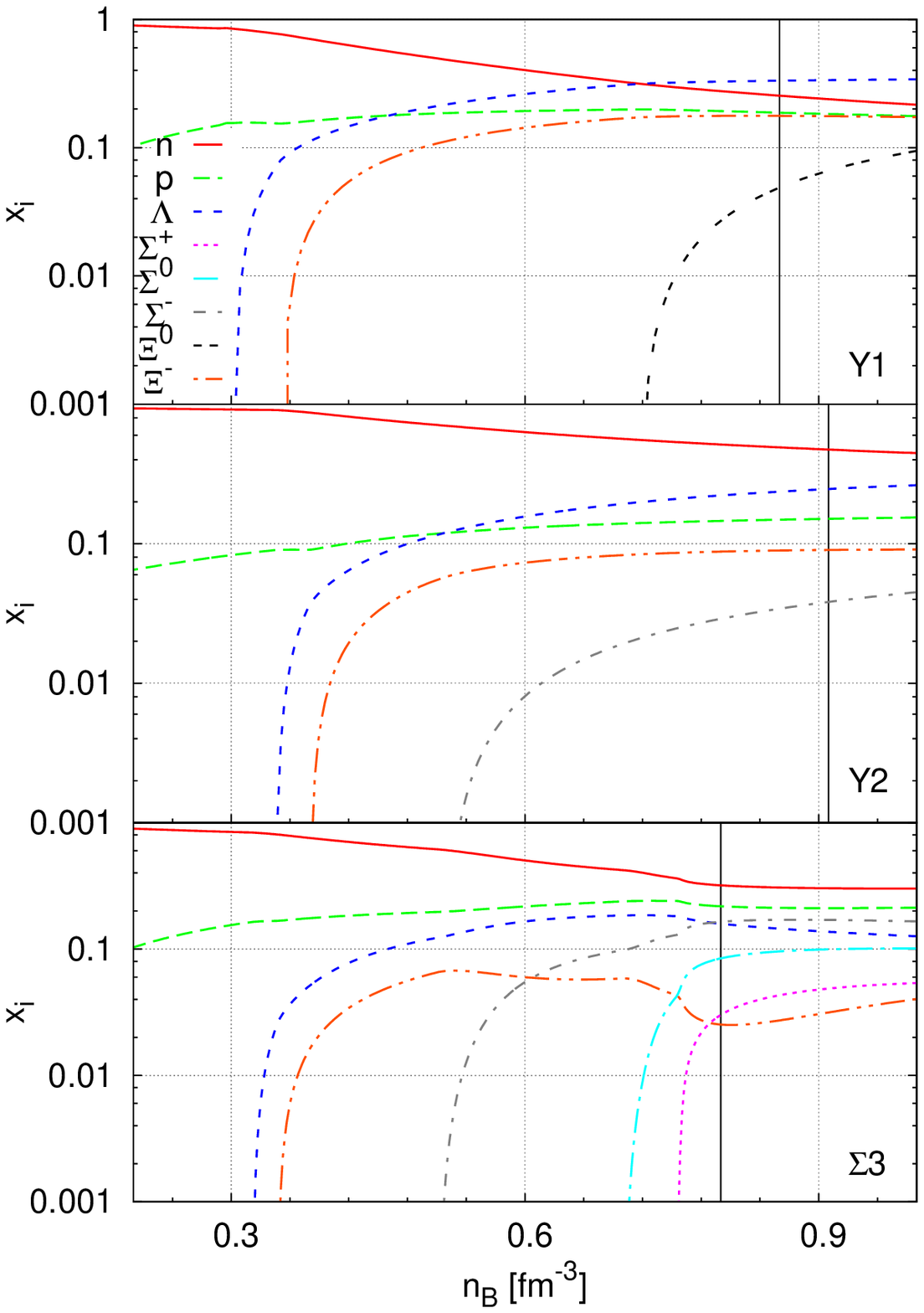}&
\includegraphics[width=0.3\linewidth,angle=0]{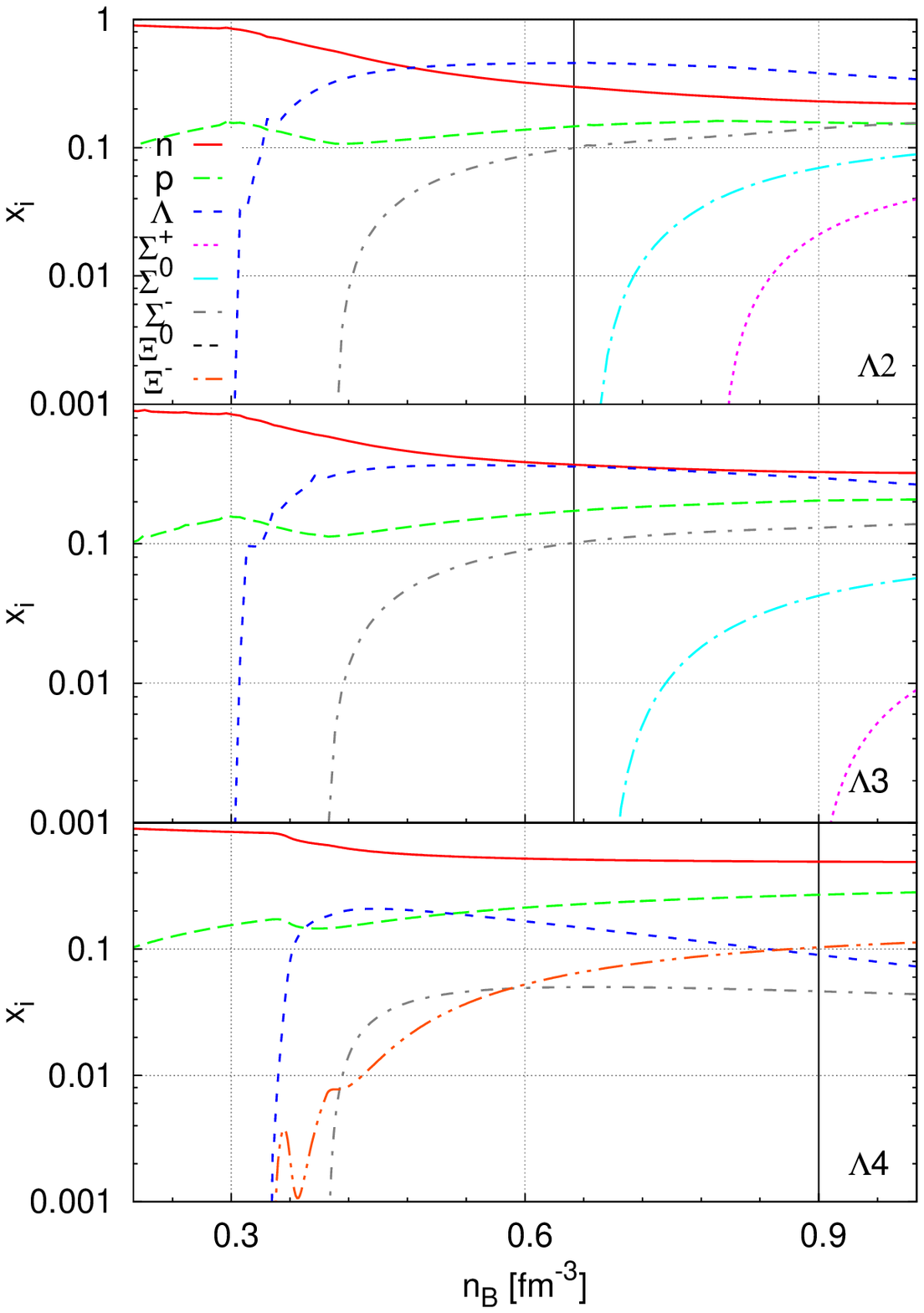}&
\includegraphics[width=0.3\linewidth,angle=0]{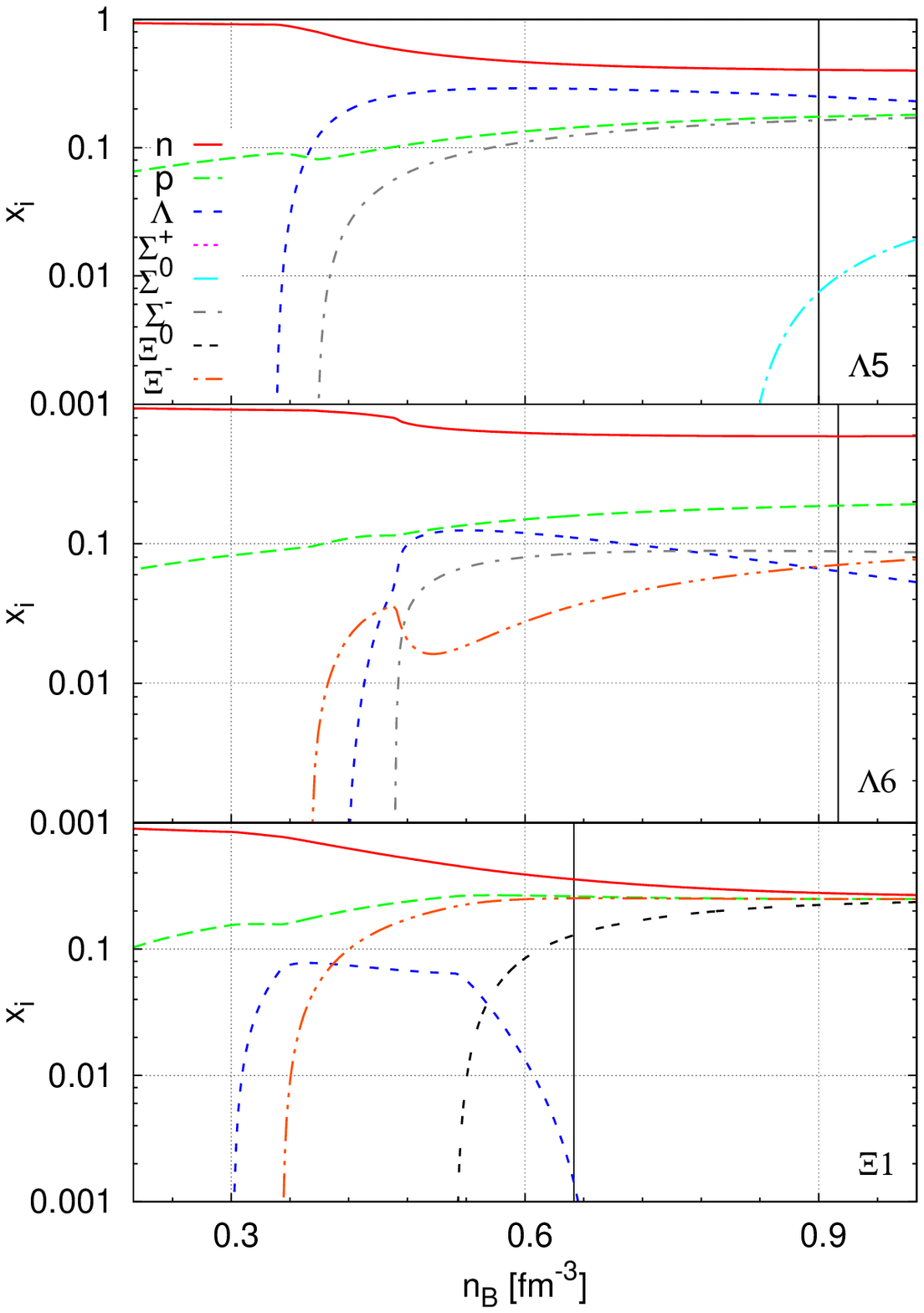}\\
\end{tabular}
\caption{\it Particle fractions in neutron star matter ($x_i = n_i/n_B$) for the
  different parameter sets discussed above within TM1-2. The vertical lines
  indicate the central density of the respective maximum mass configurations.}
\label{fracTM1}
\end{figure*}

In Fig.~\ref{fig:gm1fractions} we display the fractions of the
different particles in neutron star matter as function of baryon
number density within GM1, varying $z$ and $g_{\sigma^* i}$. As
expected, rendering the vector repulsion stronger by choosing a smaller
value of $z$, the respective hyperonic thresholds are shifted to
higher densities. Thereby, the Cascade thresholds show a stronger $z$
dependence than the $\Lambda$-threshold. The reason is the stronger
$z$-dependence of the individual hyperon-meson coupling
constants, see Eq.~(\ref{eq:symmetry}), induced by the symmetry
requirements of the procedure used. It can be observed that all
threshold densities for $z < 0.41$ are significantly lower than those
in Ref.~\cite{Weissenborn2012}, i.e., hyperons are present at much
lower baryon number densities. This can be explained by a different
readjustment of the couplings to $\sigma$: we keep the values of the
hyperon potentials in nuclear matter, $U_\Lambda^{(N)} (n_0) = -28$
MeV, $U_\Xi^{(N)} (n_0) = -18$ MeV and $U_\Sigma^{(N)} (n_0) = 30$
MeV, constant when changing the value of $z$. In addition, all the
curves have been calculated with nonzero coupling to $\sigma^*$.

The observed dependence on the attraction furnished by $\sigma^*$ is
no surprise. Increasing $g_{\sigma^* i}$, the threshold density for
hyperon $i$ is lowered and its abundance is globally increased. In
particular, on the right panel it can be seen that for a strongly
attractive coupling of $\Xi$-hyperons, although, due to the much
higher mass of the $\Xi$ with respect to $\Lambda$, the latter
threshold still remains the lowest one, at densities beyond the
$\Xi^0$-threshold, $\Lambda$-hyperons become the less abundant
ones. This is again an example which shows to which extent the
composition of the core of neutron stars depends on the interactions
between different particles and the necessity of more experimental data
to
pin down the neutron star composition.

The fractions of particles plotted in Fig.~\ref{fracTM1} were calculated
within TM1-2 ($L=110$ MeV) and TM1-2 with the $\omega\rho$ nonlinear term
($L=55$ MeV). In case of an existing instability, it is driven by the hyperon
indicated in the name of the parametrisation, see
Table~\ref{tab:tm1cassig}. Some additional comments are in order:
\begin{itemize}
\item[a)] for the hyperon potentials chosen
and taking for the isoscalar vector mesons the $SU(6)$ couplings, the first
hyperon to set in is always the $\Lambda$ followed by the $\Xi^-$; 
\item[b)] increasing the
strength of the $g_{\phi Y}$ coupling with respect to its $SU(6)$ value
disfavors the onset of $\Xi$ due to the large strangeness of these hyperons
and the onset of $\Lambda$ is followed by the onset of $\Sigma^-$; 
\item[c)]
increasing the strength of the $g_{\omega Y}$ coupling with respect to its
$SU(6)$-value will disfavor more strongly the onset of $\Sigma$ and 
$\Lambda$, and, therefore, it may happen that the $\Xi^-$ onset density is the lowest,
mainly if $L=55$ MeV, the latter having a larger $g_{\rho i}$ favoring negatively charged hyperons; 
\item[d)] taking the smaller value of the symmetry energy slope, $L=55$ MeV,
  the $\Lambda$-hyperons set in at larger densities, and the $\Sigma$-hyperons
  at smaller densities. The total strangeness inside the maximum mass star is
  smaller for the EoS with smaller $L$.
\end{itemize}

\section{Summary and conclusions}
\label{sec:summary}
We have investigated neutron star matter including hyperonic degrees of
freedom within an RMF approach.  For the nucleonic EOS we have considered the
GM1 parametrization \cite{gm91}, the DDH$\delta$ \cite{Gaitanos}, and some
variations of the TM1 parametrization \cite{tm1} with a smaller symmetry
energy slope $L$ and/or a harder EoS at large densities
\cite{providencia13}. The hyperon-nucleon interactions have been adjusted to
existing experimental data. Thereby we have followed two different strategies
in the isoscalar vector sector: either symmetry
constraints~\cite{Schaffner1996} have been imposed, relaxing the
$SU(6)$-symmetry to fix the couplings as done in several recent
works~\cite{Weissenborn2012,Miyatsu2013,Lopes2014} or no particular symmetry
has been assumed between hyperonic and nuclear couplings.  For the
hyperon-hyperon interaction, that in the present formalism is described
through the mesons with hidden strangeness, $\sigma^*$ and $\phi$, the
couplings $g_{\sigma^* i}$ have been varied freely, and, in particular, they
have been chosen strong enough to originate an instability with the onset of
hyperons; whereas for the coupling to $\phi$, the prescription for the
isoscalar vector sector has been followed.

Our main focus has been to study the possibility that an instability
driven by the onset of hyperons could exist, the neutron star maximum
mass, and the strangeness content of neutron star matter. We have
looked at the radii of intermediate mass neutron stars with EoS
containing hyperons, too. The existence of an instability as trace of
a first order phase transition was identified by analyzing the
curvature of the thermodynamical potential with respect to the
baryonic, strangeness and leptonic densities.  In all our studies at
most one negative eigenvalue, corresponding to the direction in
density space, in which density fluctuations get spontaneously and
exponentially amplified in order to achieve phase separation, has been
found.

First we have studied $n, p, e,$ and $\Lambda$ matter in $\beta$-equilibrium,
and showed that it was possible to choose a set of parameters that gives rise
to an instability driven by the onset of $\Lambda$s and still predict a
maximum star mass of the order of 2$M_\odot$, and stars with a mass of $\sim
1.4\, M_\odot$ with a radius of $12-13$ km.  It was shown that the hyperon
content is very sensitive to the attraction furnished by a coupling to
$\sigma^*$ and that the absolute value of the hyperon content is strongly
model dependent. The $npe\Lambda$ calculations are far from excluding hyperons
from neutron stars. The price to pay for having an instability is, however, a
very strong $\Lambda$-$\Lambda$-attraction, which is in contradiction with the
actual experimental information.

In a second step we have considered the whole baryonic octet. Again, it was
shown that a particular choice of the coupling parameters $g_{\sigma^* i}$,
$g_{\phi i} $, and $g_{\omega i}$ allowed the construction of EoS giving rise
to star masses as high as 2$M_\odot$, which, in addition, predict the
occurance of instabilities at the onset of hyperons. In particular, it
was shown that it is possible to have an instability driven by the onset of
the $\Lambda$, the $\Xi$ or the $\Sigma$ hyperons depending on the choice of
the coupling parameters. The coupling parameters will also determine the
different hyperon species and the strangeness fraction occuring inside a
neutron star.  Presently, the scarce amount of experimental information on the
hyperon sector, leaves too much freedom in adjusting the interaction
parameters, to give a definite answer about the composition of neutron star
matter with hyperons.

It was also shown that the neutron star radii at intermediate masses
depend mainly on the properties of the purely nuclear EoS and
considerable differences due to the presence of hyperons can be
observed only for masses close to the respective maximum mass. The
main parameter determining the radii at intermediate masses is the
slope of the symmetry energy, $L$, as found
earlier~\cite{hor01,hor03,rafael11}. It was shown that hyperons can be
added to nuclear models with low symmetry energy and slope without
violating the neutron star maximum mass constraint, and that in this
way radii between 12-13 km can be obtained for neutron stars with the
canonical mass of $1.4 M_\odot$.  From rather general arguments it seems indeed difficult to obtain even lower radii with an EoS satisfying the maximum mass constraint~\cite{Kurkela2014}. In fact, for almost all the
parameter sets considered, the $\Lambda$-hyperons appear first, and
the hyperons are present only in stars with $M\gtrsim 1.5 \, M_\odot$,
such that the radius at $M = 1.4 M_\odot$ is determined by the nuclear
parameters. Contrary to \cite{Fortin2014} we could get hyperonic stars
described within a RMF calculation with a mass $M=2M_\odot$ or above,
and still satisfying the semi-empirical constraint on the pressure of
neutron star matter at saturation density~\cite{hebeler2013}.

A selection of the EoS presented in this paper is publicly available on the
Compose web site~\cite{Typel2013}, {\it http://compose.obspm.fr}.

\acknowledgments 

This work has been partially funded by the SN2NS project 
ANR-10-BLAN-0503 and by Project PEst-OE/FIS/UI0405/2014 
developed under the initiative QREN financed by the UE/FEDER through the
program COMPETE/FCT,
 and it has been supported by
NewCompstar, COST Action MP1304.
Ad. R. R acknowledges partial support from the Romanian National
Authority for Scientific Research under grants 
PN-II-ID-PCE-2011-3-0092 and PN 09 37 01 05
and kind hospitality from LPC-Caen and LUTH-Meudon.

\end{document}